\documentclass[journal]{IEEEtran}

\usepackage{multicol}
\usepackage{graphicx}
\graphicspath{{Figures/}}
\usepackage{epstopdf}
\usepackage{acronym}
\usepackage{amsmath}
\usepackage{amsfonts}
\usepackage{color, colortbl}
\usepackage{subfigure}
\usepackage[english]{babel}
\usepackage{blindtext}
\usepackage{algorithm} 
\usepackage{algorithmic} 
\usepackage{multirow}
\usepackage{color}
\usepackage{subfigure}
\usepackage{balance}
\usepackage{caption}

\begin{document}

\newacro{3GPP}{third generation partnership project}
\newacro{4G}{4{th} generation}
\newacro{5G}{5{th} generation}

\newacro{Adam}{adaptive moment estimation}
\newacro{ADC}{analogue-to-digital converter}
\newacro{AED}{accumulated euclidean distance}
\newacro{AGC}{automatic gain control}
\newacro{AI}{artificial intelligence}
\newacro{AMB}{adaptive multi-band}
\newacro{AMB-SEFDM}{adaptive multi-band SEFDM}
\newacro{AN}{artificial noise}
\newacro{ANN}{artificial neural network}
\newacro{AoA}{angle of arrival}
\newacro{ASE}{amplified spontaneous emission}
\newacro{ASIC}{application specific integrated circuit}
\newacro{AWG}{arbitrary waveform generator}
\newacro{AWGN}{additive white Gaussian noise}
\newacro{A/D}{analog-to-digital}

\newacro{B2B}{back-to-back}
\newacro{BCF}{bandwidth compression factor}
\newacro{BCJR}{Bahl-Cocke-Jelinek-Raviv}
\newacro{BDM}{bit division multiplexing}
\newacro{BED}{block efficient detector}
\newacro{BER}{bit error rate}
\newacro{Block-SEFDM}{block-spectrally efficient frequency division multiplexing}
\newacro{BLER}{block error rate}
\newacro{BPSK}{binary phase shift keying}
\newacro{BS}{base station}
\newacro{BSS}{best solution selector}
\newacro{BT}{British Telecom}
\newacro{BU}{butterfly unit}

\newacro{CapEx}{capital expenditure}
\newacro{CA}{carrier aggregation}
\newacro{CBS}{central base station}
\newacro{CC}{component carriers}
\newacro{CCDF}{complementary cumulative distribution function}
\newacro{CCE}{control channel element}
\newacro{CCs}{component carriers}
\newacro{CD}{chromatic dispersion}
\newacro{CDF}{cumulative distribution function}
\newacro{CDI}{channel distortion information}
\newacro{CDMA}{code division multiple access}
\newacro{CI}{constructive interference}
\newacro{CIR}{carrier-to-interference power ratio}
\newacro{CMOS}{complementary metal-oxide-semiconductor}
\newacro{CNN}{convolutional neural network}
\newacro{CoMP}{coordinated multiple point}
\newacro{CO-SEFDM}{coherent optical-SEFDM}
\newacro{CP}{cyclic prefix}
\newacro{CPE}{common phase error}
\newacro{CRVD}{conventional real valued decomposition}
\newacro{CR}{cognitive radio}
\newacro{CRC}{cyclic redundancy check}
\newacro{CS}{central station}
\newacro{CSI}{channel state information}
\newacro{CSIT}{channel state information at the transmitter}
\newacro{CSPR}{carrier to signal power ratio}
\newacro{CW}{continuous-wave}
\newacro{CWT}{continuous wavelet transform}
\newacro{C-RAN}{cloud-radio access networks}

\newacro{DAC}{digital-to-analogue converter}
\newacro{DBP}{digital backward propagation}
\newacro{DC}{direct current}
\newacro{DCGAN}{deep convolutional generative adversarial network}
\newacro{DCI}{downlink control information}
\newacro{DCT}{discrete cosine transform}
\newacro{DDC}{digital down-conversion}
\newacro{DDO-OFDM}{directed detection optical-OFDM}
\newacro{DDO-OFDM}{direct detection optical-OFDM}
\newacro{DDO-SEFDM}{directed detection optical-SEFDM}
\newacro{DFB}{distributed feedback}
\newacro{DFDMA}{distributed FDMA}
\newacro{DFT}{discrete Fourier transform}
\newacro{DFrFT}{discrete fractional Fourier transform}
\newacro{DL}{deep learning}
\newacro{DMA}{direct memory access}
\newacro{DMRS}{demodulation reference signal}
\newacro{DoF}{degree of freedom}
\newacro{DOFDM}{dense orthogonal frequency division multiplexing}
\newacro{DP}{dual polarization}
\newacro{DPC}{dirty paper coding}
\newacro{DSB}{double sideband}
\newacro{DSL}{digital subscriber line}
\newacro{DSP}{digital signal processors}
\newacro{DSSS}{direct sequence spread spectrum}
\newacro{DT}{decision tree}
\newacro{DVB}{digital video broadcast}
\newacro{DWDM}{dense wavelength division multiplexing}
\newacro{DWT}{discrete wavelet transform}
\newacro{D/A}{digital-to-analog}

\newacro{ECC}{error correcting codes}
\newacro{ECL}{external-cavity laser}
\newacro{ECOC}{error-correcting output codes}
\newacro{EDFA}{erbium doped fiber amplifier}
\newacro{EE}{energy efficiency}
\newacro{eMBB}{enhanced mobile broadband}
\newacro{eNB-IoT}{enhanced NB-IoT}
\newacro{EPA}{extended pedestrian A}
\newacro{EVM}{error vector magnitude}

\newacro{Fast-OFDM}{fast-orthogonal frequency division multiplexing}
\newacro{FBMC}{filter bank multicarrier }
\newacro{FCE}{full channel estimation}
\newacro{FD}{fixed detector}
\newacro{FDD}{frequency division duplexing}
\newacro{FDM}{frequency division multiplexing}
\newacro{FDMA}{frequency division multiple access}
\newacro{FE}{full expansion}
\newacro{FEC}{forward error correction}
\newacro{FEXT}{far-end crosstalk}
\newacro{FF}{flip-flop}
\newacro{FFT}{fast Fourier transform}
\newacro{FFTW}{Fastest Fourier Transform in the West}
\newacro{FHSS}{frequency-hopping spread spectrum}
\newacro{FIFO}{first in first out}
\newacro{FMCW}{frequency-modulated continuous wave}
\newacro{F-OFDM}{filtered-orthogonal frequency division multiplexing}
\newacro{FPGA}{field programmable gate array}
\newacro{FrFT}{fractional Fourier transform}
\newacro{FSD}{fixed sphere decoding}
\newacro{FSD-MNSF}{FSD-modified-non-sort-free}
\newacro{FSD-NSF}{FSD-non-sort-free}
\newacro{FSD-SF}{FSD-sort-free}
\newacro{FSK}{frequency shift keying}
\newacro{FTN}{faster than Nyquist}
\newacro{FTTB}{fiber to the building}
\newacro{FTTC}{fiber to the cabinet}
\newacro{FTTdp}{fiber to the distribution point}
\newacro{FTTH}{fiber to the home}

\newacro{GAN}{generative adversarial network}
\newacro{GB}{guard band}
\newacro{GFDM}{generalized frequency division multiplexing}
\newacro{GNN}{graph neural networks}
\newacro{GPU}{graphics processing unit}
\newacro{GSM}{global system for mobile communication}
\newacro{GUI}{graphical user interface}

\newacro{HARQ}{hybrid automatic repeat request}
\newacro{HC-MCM}{high compaction multi-carrier communication}
\newacro{HPA}{high power amplifier}

\newacro{IC}{integrated circuit}
\newacro{ICI}{inter carrier interference}
\newacro{ICT}{Information and communications technology}
\newacro{ID}{iterative detection}
\newacro{IDCT}{inverse discrete cosine transform}
\newacro{IDFT}{inverse discrete Fourier transform}
\newacro{IDFrFT}{inverse discrete fractional Fourier transform}
\newacro{ID-FSD}{iterative detection-FSD}
\newacro{ID-SD}{ID-sphere decoding}
\newacro{IF}{intermediate frequency}
\newacro{IFFT}{inverse fast Fourier transform}
\newacro{IFrFT}{inverse fractional Fourier transform}
\newacro{IIoT}{industrial Internet of things}
\newacro{IM}{index modulation}
\newacro{IMD}{intermodulation distortion}
\newacro{INOFS}{inverse non-orthogonal frequency shaping}
\newacro{IoT}{Internet of things}
\newacro{IOTA}{isotropic orthogonal transform algorithm}
\newacro{IP}{intellectual property}
\newacro{IR}{infrared}
\newacro{ISAC}{integrated sensing and communication}
\newacro{ISAR}{inverse synthetic aperture radar}
\newacro{ISC}{interference self cancellation}
\newacro{ISI}{inter symbol interference}
\newacro{ISM}{industrial, scientific and medical}
\newacro{ISTA}{iterative shrinkage and thresholding}
\newacro{ITU}{international telecommunication union}
\newacro{IUI}{inter user interference}
\newacro{IWAI}{integrated waveform and intelligence}

\newacro{KNN}{k-nearest neighbours}

\newacro{LDPC}{low density parity check}
\newacro{LFDMA}{localized FDMA}
\newacro{LLR}{log-likelihood ratio}
\newacro{LNA}{low noise amplifier}
\newacro{LO}{local oscillator}
\newacro{LOS}{line-of-sight}
\newacro{LPWAN}{low power wide area network}
\newacro{LS}{least square}
\newacro{LSTM}{long short-term memory}
\newacro{LTE}{long term evolution}
\newacro{LTE-Advanced}{long term evolution-advanced}
\newacro{LUT}{look-up table}

\newacro{MA}{multiple access}
\newacro{MAC}{media access control}
\newacro{MAMB}{mixed adaptive multi-band}
\newacro{MAMB-SEFDM}{mixed adaptive multi-band SEFDM}
\newacro{MASK}{m-ary amplitude shift keying}
\newacro{MB}{multi-band}
\newacro{MB-SEFDM}{multi-band SEFDM}
\newacro{MCM}{multi-carrier modulation}
\newacro{MC-CDMA}{multi-carrier code division multiple access}
\newacro{MCS}{modulation and coding scheme}
\newacro{MF}{matched filter}
\newacro{MIMO}{multiple input multiple output}
\newacro{ML}{maximum likelihood}
\newacro{MLSD}{maximum likelihood sequence detection}
\newacro{MMF}{multi-mode fiber}
\newacro{MMSE}{minimum mean squared error}
\newacro{mMTC}{massive machine-type communication}
\newacro{MNSF}{modified-non-sort-free}
\newacro{MOFDM}{masked-OFDM}
\newacro{MRVD}{modified real valued decomposition}
\newacro{MS}{mobile station}
\newacro{MSE}{mean squared error}
\newacro{MTC}{machine-type communication}
\newacro{MUI}{multi-user interference}
\newacro{MUSA}{multi-user shared access}
\newacro{MU-MIMO}{multi-user multiple-input multiple-output}
\newacro{MZM}{Mach-Zehnder modulator}
\newacro{M2M}{machine to machine}

\newacro{NB-IoT}{narrowband IoT}
\newacro{NB}{naive Bayesian}
\newacro{NDFF}{National Dark Fiber Facility}
\newacro{NEXT}{near-end crosstalk}
\newacro{NFV}{network function virtualization}
\newacro{NG-IoT}{next generation IoT}
\newacro{NLOS}{non-line-of-sight}
\newacro{NLSE}{nonlinear Schrödinger equation}
\newacro{NN}{neural network}
\newacro{NOFDM}{non-orthogonal frequency division multiplexing}
\newacro{NOMA}{non-orthogonal multiple access}
\newacro{NoFDMA}{non-orthogonal frequency division multiple access}
\newacro{NOFS}{non-orthogonal frequency spacing}
\newacro{NP}{non-polynomial}
\newacro{NR}{new radio}
\newacro{NSF}{non-sort-free}
\newacro{NWDM}{Nyquist wavelength division multiplexing }
\newacro{Nyquist-SEFDM}{Nyquist-spectrally efficient frequency division multiplexing}

\newacro{OBM-OFDM}{orthogonal band multiplexed OFDM}
\newacro{ODDM}{orthogonal delay-Doppler division multiplexing}
\newacro{OF}{optical filter}
\newacro{OFDM}{orthogonal frequency division multiplexing}
\newacro{OFDMA}{orthogonal frequency division multiple access}
\newacro{OMA}{orthogonal multiple access}
\newacro{OpEx}{operating expenditure}
\newacro{OPM}{optical performance monitoring}
\newacro{OQAM}{offset-QAM}
\newacro{OSI}{open systems interconnection}
\newacro{OSNR}{optical signal-to-noise ratio}
\newacro{OSSB}{optical single sideband}
\newacro{OTA}{over-the-air}
\newacro{OTFS}{orthogonal time frequency space}
\newacro{Ov-FDM}{Overlapped FDM}
\newacro{O-SEFDM}{optical-spectrally efficient frequency division multiplexing}
\newacro{O-FOFDM}{optical-fast orthogonal frequency division multiplexing}
\newacro{O-OFDM}{optical-orthogonal frequency division multiplexing}
\newacro{O-CDMA}{optical-code division multiple access}

\newacro{PA}{power amplifier}
\newacro{PAPR}{peak-to-average power ratio}
\newacro{PCA}{principal component analysis}
\newacro{PCE}{partial channel estimation}
\newacro{PD}{photodiode}
\newacro{PDCCH}{physical downlink control channel}
\newacro{PDF}{probability density function}
\newacro{PDP}{power delay profile}
\newacro{PDMA}{polarisation division multiple access}
\newacro{PDM-OFDM}{polarization-division multiplexing-OFDM}
\newacro{PDM-SEFDM}{polarization-division multiplexing-SEFDM}
\newacro{PDSCH}{physical downlink shared channel}
\newacro{PE}{processing element}
\newacro{PED}{partial Euclidean distance}
\newacro{PLA}{physical layer authentication}
\newacro{PLS}{physical layer security}
\newacro{PMD}{polarization mode dispersion}
\newacro{PON}{passive optical network}
\newacro{PPM}{parts per million}
\newacro{PRB}{physical resource block}
\newacro{PSD}{power spectral density}
\newacro{PSK}{pre-shared key}
\newacro{PSNR}{peak signal-to-noise ratio}
\newacro{PSS}{primary synchronization signal}
\newacro{PU}{primary user}
\newacro{PXI}{PCI extensions for instrumentation}
\newacro{P/S}{parallel-to-serial}

\newacro{QAM}{quadrature amplitude modulation}
\newacro{QKD}{quantum key distribution}
\newacro{QoS}{quality of service}
\newacro{QPSK}{quadrature phase-shift keying}
\newacro{QRNG}{quantum random number generation}

\newacro{RAUs}{remote antenna units}
\newacro{RAT}{radio access technology}
\newacro{RBF}{radial basis function}
\newacro{RBW}{resolution bandwidth}
\newacro{ReLU}{rectified linear units}
\newacro{RF}{radio frequency}
\newacro{RMS}{root mean square}
\newacro{RMSE}{root mean square error}
\newacro{RMSProp}{root mean square propagation}
\newacro{RNTI}{radio network temporary identifier}
\newacro{RoF}{radio-over-fiber}
\newacro{ROM}{read only memory}
\newacro{RRC}{root raised cosine}
\newacro{RC}{raised cosine}
\newacro{RSC}{recursive systematic convolutional}
\newacro{RSSI}{received signal strength indicator}
\newacro{RTL}{register transfer level}
\newacro{RVD}{real valued decomposition}

\newacro{SB-SEFDM}{single-band SEFDM}
\newacro{ScIR}{sub-carrier to interference ratio}
\newacro{SCMA}{sparse code multiple access}
\newacro{SC-NOFS}{single-carrier non-orthogonal frequency shaping}
\newacro{SC-OFDM}{single-carrier orthogonal frequency division multiplexing}
\newacro{SC-FDMA}{single-carrier frequency division multiple access}
\newacro{SC-SEFDMA}{single-carrier spectrally efficient frequency division multiple access}
\newacro{SD}{sphere decoding}
\newacro{SDM}{space division multiplexing}
\newacro{SDMA}{space division multiple access}
\newacro{SDN}{software-defined network}
\newacro{SDP}{semidefinite programming}
\newacro{SDR}{software-defined radio}
\newacro{SE}{spectral efficiency}
\newacro{SEFDM}{spectrally efficient frequency division multiplexing}
\newacro{SEFDMA}{spectrally efficient frequency division multiple access} 
\newacro{SF}{sort-free}
\newacro{SFCW}{stepped-frequency continuous wave}
\newacro{SGD}{stochastic gradient descent}
\newacro{SGDM}{stochastic gradient descent with momentum}
\newacro{SIC}{successive interference cancellation}
\newacro{SiGe}{silicon-germanium}
\newacro{SINR}{signal-to-interference-plus-noise ratio}
\newacro{SIR}{signal-to-interference ratio}
\newacro{SISO}{single-input single-output}
\newacro{SLM}{spatial light modulator}
\newacro{SMF}{single mode fiber}
\newacro{SNR}{signal-to-noise ratio}
\newacro{SP}{shortest-path}
\newacro{SPSC}{symbol per signal class}
\newacro{SPM}{self-phase modulation}
\newacro{SRS}{sounding reference signal}
\newacro{SSB}{single-sideband}
\newacro{SSBI}{signal-signal beat interference}
\newacro{SSFM}{split-step Fourier method}
\newacro{SSMF}{standard single mode fiber}
\newacro{STBC}{space time block coding}
\newacro{STFT}{short time Fourier transform}
\newacro{STC}{space time coding}
\newacro{STO}{symbol timing offset}
\newacro{SU}{secondary user}
\newacro{SVD}{singular value decomposition}
\newacro{SVM}{support vector machine}
\newacro{SVR}{singular value reconstruction}
\newacro{S/P}{serial-to-parallel}

\newacro{TDD}{time division duplexing}
\newacro{TDMA}{time division multiple access }
\newacro{TDM}{time division multiplexing}
\newacro{TFP}{time frequency packing}
\newacro{THP}{Tomlinson-Harashima precoding}
\newacro{TOFDM}{truncated OFDM}
\newacro{TSPSC}{training symbols per signal class}
\newacro{TSVD}{truncated singular value decomposition}
\newacro{TSVD-FSD}{truncated singular value decomposition-fixed sphere decoding}
\newacro{TTI}{transmission time interval}

\newacro{UAV}{unmanned aerial vehicle}
\newacro{UCR}{user compression ratio}
\newacro{UE}{user equipment}
\newacro{UFMC}{universal-filtered multi-carrier}
\newacro{ULA}{uniform linear array}
\newacro{UMTS}{universal mobile telecommunications service}
\newacro{URLLC}{ultra-reliable low-latency communication}
\newacro{USRP}{universal software radio peripheral}
\newacro{UWB}{ultra-wideband}

\newacro{VDSL}{very-high-bit-rate digital subscriber line}
\newacro{VDSL2}{very-high-bit-rate digital subscriber line 2}
\newacro{VHDL}{very high speed integrated circuit hardware description language}
\newacro{VLC}{visible light communication}
\newacro{VLSI}{very large scale integration}
\newacro{VOA}{variable optical attenuator}
\newacro{VP}{vector perturbation}
\newacro{VSSB-OFDM}{virtual single-sideband OFDM}
\newacro{V2V}{vehicle-to-vehicle}

\newacro{WAN}{wide area network}
\newacro{WCDMA}{wideband code division multiple access}
\newacro{WDM}{wavelength division multiplexing}
\newacro{WDP}{waveform-defined privacy}
\newacro{WDS}{waveform-defined security}
\newacro{WiFi}{wireless fidelity}
\newacro{WiGig}{Wireless Gigabit Alliance}
\newacro{WiMAX}{Worldwide interoperability for Microwave Access}
\newacro{WLAN}{wireless local area network}
\newacro{WSS}{wavelength selective switch}

\newacro{XPM}{cross-phase modulation}

\newacro{ZF}{zero forcing}
\newacro{ZP}{zero padding}


\title{A Low-Cost Multi-Band Waveform Security Framework in Resource-Constrained Communications}

\author{{Tongyang Xu,~\IEEEmembership{Member,~IEEE}, Zhongxiang Wei,~\IEEEmembership{Member,~IEEE}, Tianhua Xu,~\IEEEmembership{Member,~IEEE} and \\ Gan Zheng,~\IEEEmembership{Fellow,~IEEE}}
\thanks{
This work was supported in part by the UK Engineering and Physical Sciences Research Council (EPSRC) under Grant EP/Y000315/1 and Grant EP/X04047X/1, in part by the Natural Science Foundation of China under Grants 62101384, in part by the Guangdong Basic and Applied Basic Research Foundation under Grant 2022B1515120018, in part by European Union (EU) Horizon 2020 Marie Sklodowska-Curie Actions (MSCA) under Grant 101008280 (DIOR) and EU Horizon Europe under Grant 101131146 (UPGRADE), and in part by the UK Royal Society under Grant (IES/R3/223068). \emph{(Corresponding authors: Tongyang Xu, Zhongxiang Wei.)}

T. Xu is with the School of Engineering, Newcastle University, Newcastle upon Tyne, NE1 7RU, U.K.  (e-mail: tongyang.xu@newcastle.ac.uk). 

Z. Wei is with the School of Electronics and Information Engineering, Tongji University, China (e-mail: z\_wei@tongji.edu.cn). 

T. Xu is with the School of Engineering, University of Warwick, Coventry CV4 7AL, United Kingdom (e-mail:tianhua.xu@warwick.ac.uk).

G. Zheng is with the School of Engineering, University of Warwick, Coventry CV4 7AL, United Kingdom (e-mail:gan.zheng@warwick.ac.uk).

}}

\maketitle

\begin{abstract}

Traditional physical layer secure beamforming is achieved via precoding before signal transmission using channel state information (CSI). However, imperfect CSI will compromise the performance with imperfect beamforming and potential information leakage. In addition, multiple RF chains and antennas are needed to support the narrow beam generation, which complicates hardware implementation and is not suitable for resource-constrained Internet-of-Things (IoT) devices. Moreover, with the advancement of hardware and artificial intelligence (AI), low-cost and intelligent eavesdropping to wireless communications is becoming increasingly detrimental. In this paper, we propose a multi-carrier based multi-band waveform-defined security (WDS) framework, independent from CSI and RF chains, to defend against AI eavesdropping. Ideally, the continuous variations of sub-band structures lead to an infinite number of spectral features, which can potentially prevent brute-force eavesdropping. Sub-band spectral pattern information is efficiently constructed at legitimate users via a proposed chaotic sequence generator. A novel security metric, termed signal classification accuracy (SCA), is used to evaluate the security robustness under AI eavesdropping. Communication error probability and complexity are also investigated to show the reliability and practical capability of the proposed framework. Finally, compared to traditional secure beamforming techniques, the proposed multi-band WDS framework reduces power consumption by up to six times.

\end{abstract}


\section{Introduction} \label{sec:intro}

\IEEEPARstart{C}{ommunication} security is an increasingly important research topic with the commercialization of 5G and the rapid development of its beyond. In typical \ac{RF} based communications, due to the broadcast nature of wireless channels, legitimate user communications are vulnerable to eavesdropping. In traditional eavesdropping scenarios, physical layer secure beamforming \cite{adversarial_survey_2016, adversarial_survey_2019, adversarial_JASC_2018} is a commonly used \ac{PLS} technique, which can prevent eavesdroppers from intercepting confidential data via optimizing spatial signal beams according to channel conditions. However, the secure beamforming techniques are showing limitations \cite{PLS_challenges_CM_2015, VLC_security_Network_survey_2022, PLS_metasurface_2022}. {Firstly, confidential data is vulnerable in the presence of multi-antenna eavesdroppers or distributed eavesdroppers and extra processing complexity is required to mitigate the challenges \cite{VLC_beamforming_TWC_2018, disEve_NatureE_2021}.} Experiments in \cite{PLS_practical_CNS_mmWave_2015} even revealed that an eavesdropper can capture confidential data from directional millimeter waves via using small-scale reflection objects. Moreover, existing security techniques require additional hardware complexity in utilizing multiple antennas and multiple RF chains, which are energy inefficient and  against net zero sustainable development objectives \cite{GreenCom_TGCN_2022}. Therefore, the high energy consumption from extra hardware utilization prevents the use of secure beamforming in low-cost \ac{IoT} applications \cite{PLS_CI_IoT_CM_2020, IoT_security_Proceeding_2015, security_IoT_TIFS_2018, JSAC_2015_zhongxiang}. More importantly, traditional secure beamforming techniques require the knowledge of \ac{CSI}. However, \ac{CSI} could be inaccurate \cite{Eavesdropping_imperfect_CSI_TWC_2012} due to pilot spoofing attacks, pilot contamination, and pilot jamming. Therefore, extra processing complexity is required to mitigate the challenges \cite{pilot_contamination_2022, pilot_spoofing_2023}. However, the acquisition of \ac{CSI} is becoming more costly \cite{GreenCom_CM_2020} especially for resource and power limited IoT applications.

Due to the advancement of \ac{AI}, a passive eavesdropper could become an active attacker resulting in \ac{AI} based threats to communication security. As an attacker, adversarial machine learning \cite{adversarial_attack_WCL_2019, adversarial_examples_2017, Adversarial_TIFS_2020} can intelligently eavesdrop and further manipulate legitimate user signal characteristics over the air, which could cause signal processing failure at a legitimate user. The adversarial attack challenges end-to-end autoencoder deep learning systems in \cite{adversarial_attack_CL_2019}, \ac{OFDM} channel estimation and signal detection in \cite{adversarial_arXiv_2021robust}, \ac{MIMO} channel estimation in \cite{adversarial_MIMO_JCN_2020}, deep learning MIMO power allocation in \cite{adversarial_MIMO_ICC_2021} and cooperative spectrum sensing in \cite{adversarial_luo_2020}. A more detrimental type of attack is termed \ac{GAN} \cite{GAN_NIPS2014}, which can simultaneously learn legitimate user signal patterns and channel/hardware impairment models to starve scarce over-the-air resources \cite{GAN_TCCN_2020} via spoofing attacks. Existing countermeasures for adversarial machine learning attacks is either sending fake data and labels to fool adversaries \cite{adversarial_attack_ICC_2018} or proactively applying adversarial attacks to intruders to prevent signal detections \cite{Adversarial_defense_2021}. However, the above methods would reduce spectral efficiency and increase system complexity.

The motivation of this work is to prevent \ac{AI} based eavesdropping and subsequent AI attacks, especially for resource-constrained secure communication scenarios, from a waveform design perspective. Typical 4G/5G systems employ the \ac{OFDM} waveform \cite{Erik_book_4G,Erik_book_5G}, which is simple for signal generation and detection but at the cost of security vulnerability. Further investigations on waveform security lead to the study of new waveform design. {There are some existing research works on designing waveforms in physical layer security. Masked-OFDM \cite{MOFDM_PIMRC2009} combines two \ac{OFDM} signals with overlapping to produce a composite non-orthogonal signal and therefore complicates eavesdropping signal detections. However, this approach also results in high complexity at legitimate user side signal detection. Work in \cite{PLS_FTN_ICCE_2017} employs variable time interval patterns to complicate eavesdropping. However, with the advancement of \ac{AI}, eavesdroppers could easily identify different patterns using intelligent algorithms. The recent work in \cite{Tongyang_JIOT_WDS_2022} proposed a \ac{WDS} framework, but the framework is still vulnerable to AI based eavesdropping. }

{This work focuses on optimizing \ac{WDS}, which is the initial waveform candidate proposed to defend against AI based eavesdropping. To enhance the traditional WDS scheme's robustness to AI eavesdropping, this work proposes an adaptive multi-band WDS framework aiming to further improve communication security.} {Multi-band waveform architectures can separate a single-band signal into multiple sub-bands. In this case, more spectral ambiguity will be introduced since each sub-band can have independent and unique spectral features. The enhanced spectral ambiguity will prevent \ac{AI} based eavesdropping and therefore avoid adversarial attacks.} {It is noted that this work aims for single user scenarios where a user occupies all sub-bands. The use of multi-band architectures is to simplify signal detection and enhance ambiguity rather than supporting multiple users using a multiple access scheme.} The fundamental principle behind \ac{WDS} is the utilization of non-orthogonal waveform \ac{SEFDM} \cite{TongyangTVT2017}, which introduces feature ambiguity via intentionally tuning sub-carrier packing patterns. As indicated by \cite{GreenCom_Nature_Electronics_2017, GreenCom_6G_summit_2020}, increased number of antennas or RF chains are the main energy consumption source. Therefore, the proposed multi-band \ac{WDS} framework, although requiring extra signal processing, can prevent AI based interception for resource-constrained IoT scenarios while available PLS techniques are too costly to implement.

The main contributions of this work are as follows:
\begin{itemize} 

\item{ A multi-band WDS secure communication framework is proposed for over-the-air \ac{PLS} scenarios aiming to defend against AI eavesdropping. {Typically, coding can encrypt signals but it cannot prevent AI eavesdropping and the variations of coding rates will complicate signal frame design and hardware implementation.} Unlike traditional beamforming \ac{PLS} approaches that require multiple antennas, the proposed framework is able to enhance \ac{PLS} security for single-antenna transceivers, which is particular suitable to resource-constrained \ac{IoT} applications. Sub-carriers are packed non-orthogonally and the packing schemes are adaptively adjustable in each sub-band, thus significantly complicating eavesdropping signal detection. {Ideally, the continuous variations of sub-band spectral compression features further enhance the \ac{PLS} by introducing an infinite number of signal patterns}, which prevents accurate signal identifications at the eavesdropper and is robust to exhaustive brute-force eavesdropping. Therefore, the proposed multi-band WDS has further enhanced security than single-band WDS by jointly complicating eavesdropping signal detection and preventing accurate signal pattern identification.}

\item{{AI security metric, termed signal classification accuracy (SCA), is proposed to replace the traditional non-AI security metric \ac{SNR}.} The eavesdropping classification accuracy approximation model is derived for the adaptive multi-band WDS framework. It shows a perfect match between the analytical model and actual results. It also reveals that the classification accuracy will further degrade by increasing the number of signal patterns and sub-bands.}

\item{{A paired-key generator is designed to ensure fast and reliable pattern information generation at both legitimate users. Prior to the key generation, the same bifurcation parameter, initial state, chaotic mapping and pattern threshold should be pre-shared and stored at both legitimate users. Using identical parameter initializations, two identical pattern generators will continuously output identical chaotic sequences, which will be used as pattern keys to produce a correlation matrix. The key generation scheme is practical since only four parameters are needed and stored in memory in advance.}}

\item{Lower implementation complexity is achieved by the multi-band waveform security framework such that the framework is suitable for low-cost and resource-constrained communication scenarios where RF chains, antennas and carrier frequency are limited. This work also reveals that the waveform security framework can reduce power consumption by up to six times compared to traditional secure beamforming techniques indicating the suitability of the framework in net-zero communications.  }

\end{itemize}

{Notations: Unless otherwise specified, matrices are denoted by bold uppercase letters (i.e., $\mathbf{F}$), vectors are represented by bold lowercase letters (i.e., $\mathbf{x}$, $\mathbf{s}$), and scalars are denoted by normal font (i.e., $\rho$). Subscripts indicate the location of the entry in the matrices or vectors (i.e., $c_{i,j}$ and $s_n$ are the ($i,j$)-th and the $n$-th element in $\mathbf{C}$ and $\mathbf{s}$, respectively)}

\section{The Principle of WDS Framework} \label{sec:eve_classification}

\begin{figure}[t!]
\begin{center}
\includegraphics[scale=0.5]{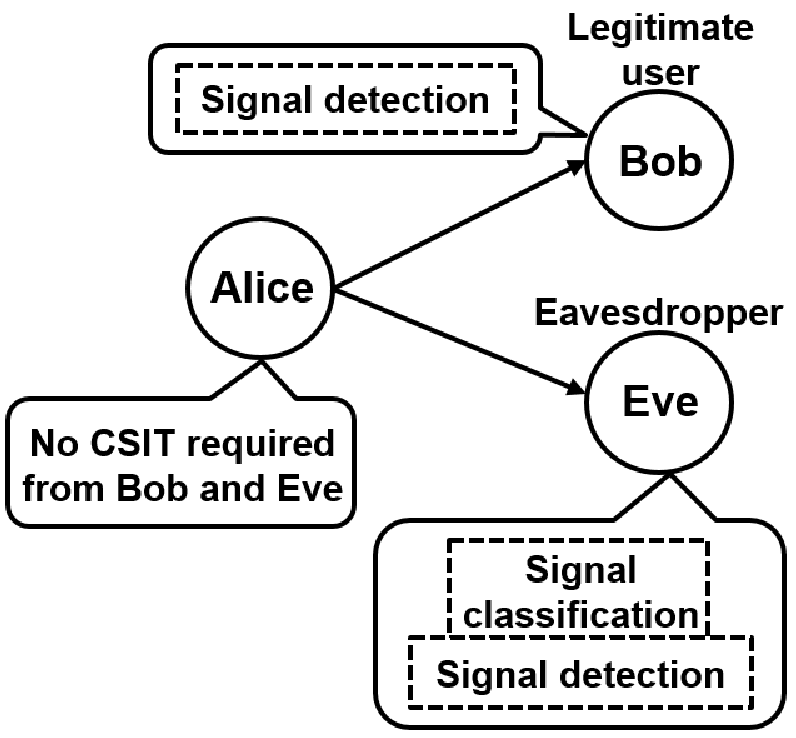}
\end{center}
\caption{The waveform based secure communication model for legitimate user and eavesdropper.}
\label{Fig:MAMB_eavesdropping_model}
\end{figure}

The principle of the waveform-defined security communication framework is demonstrated in Fig. \ref{Fig:MAMB_eavesdropping_model}. Traditional \ac{PLS} techniques aim to weaken the wiretap link while enhancing the legitimate link using beamforming. However, they require \ac{CSIT} from both eavesdroppers and legitimate users, which are commonly unavailable in most cases. The proposed WDS framework avoids \ac{CSIT} and therefore simplifies the entire system design. Unlike traditional multi-antenna based beamforming defence techniques, a WDS communication system will employ an omni-directional communication format using a single antenna. In this case, the WDS framework saves antennas and RF chains leading to reduced hardware complexity. It is noted that the WDS framework is also applicable in multi-antenna systems, in which it can enhance the over-the-air encryption of beamforming. The eavesdropper is assumed to be passive in this work, therefore it will firstly learn to identify signal patterns and then detect signals. In this case, the aim of WDS is to design signal patterns that will prevent accurate signal classification and complicate signal detection at eavesdroppers.

\subsection{Signal Pattern Principle} \label{subsec:waveform_principle}

\begin{figure}[t!]
\begin{center}
\includegraphics[scale=0.41]{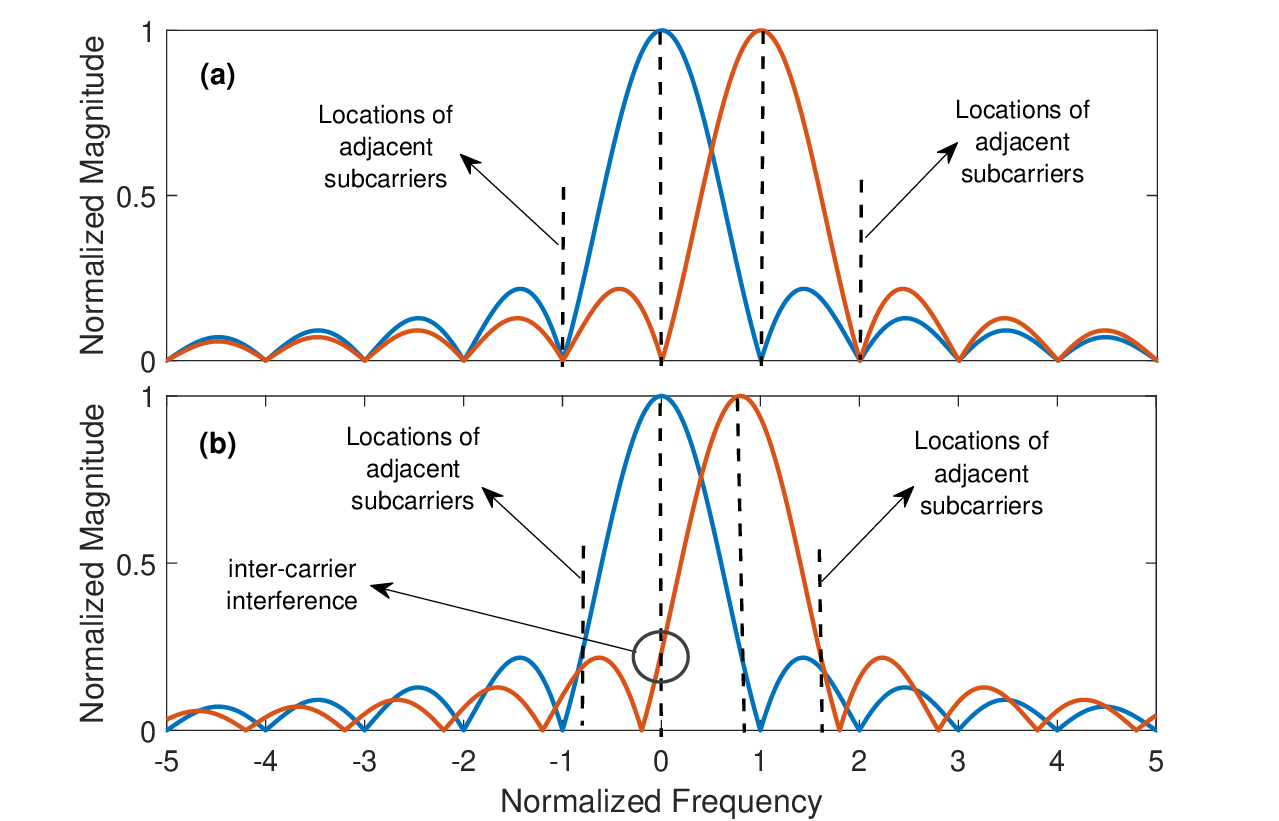}
\end{center}
\caption{Principle of non-orthogonal SEFDM signal waveform. (a) OFDM sub-carrier packing. (b) SEFDM sub-carrier packing.}
\label{Fig:AI_encryption_subcarrier_packing}
\end{figure}

The traditional OFDM is a multi-carrier signal with sub-carrier spacing of $\Delta{f}=1/{T}$ where $T$ is the time duration of one OFDM symbol. The principle of SEFDM is to pack sub-carriers closer in a non-orthogonal format while maintaining the bandwidth for each sub-carrier. Therefore, the sub-carrier spacing becomes $\Delta{f}=\alpha/{T}$ where $\alpha<$1 is the \ac{BCF}, which determines the bandwidth compression ratio. The spectral bandwidth compression principle for SEFDM is illustrated in Fig. \ref{Fig:AI_encryption_subcarrier_packing} {(reused from \cite{Tongyang_JIOT_WDS_2022})} where the spectral efficiency improvement of SEFDM over OFDM is given by
\begin{equation}
\eta=(\frac{1}{\alpha}-1)\times{100}.
\label{eq:SEFDM_SE_increase}\end{equation}

\begin{figure}[t!]
\begin{center}
\includegraphics[scale=0.38]{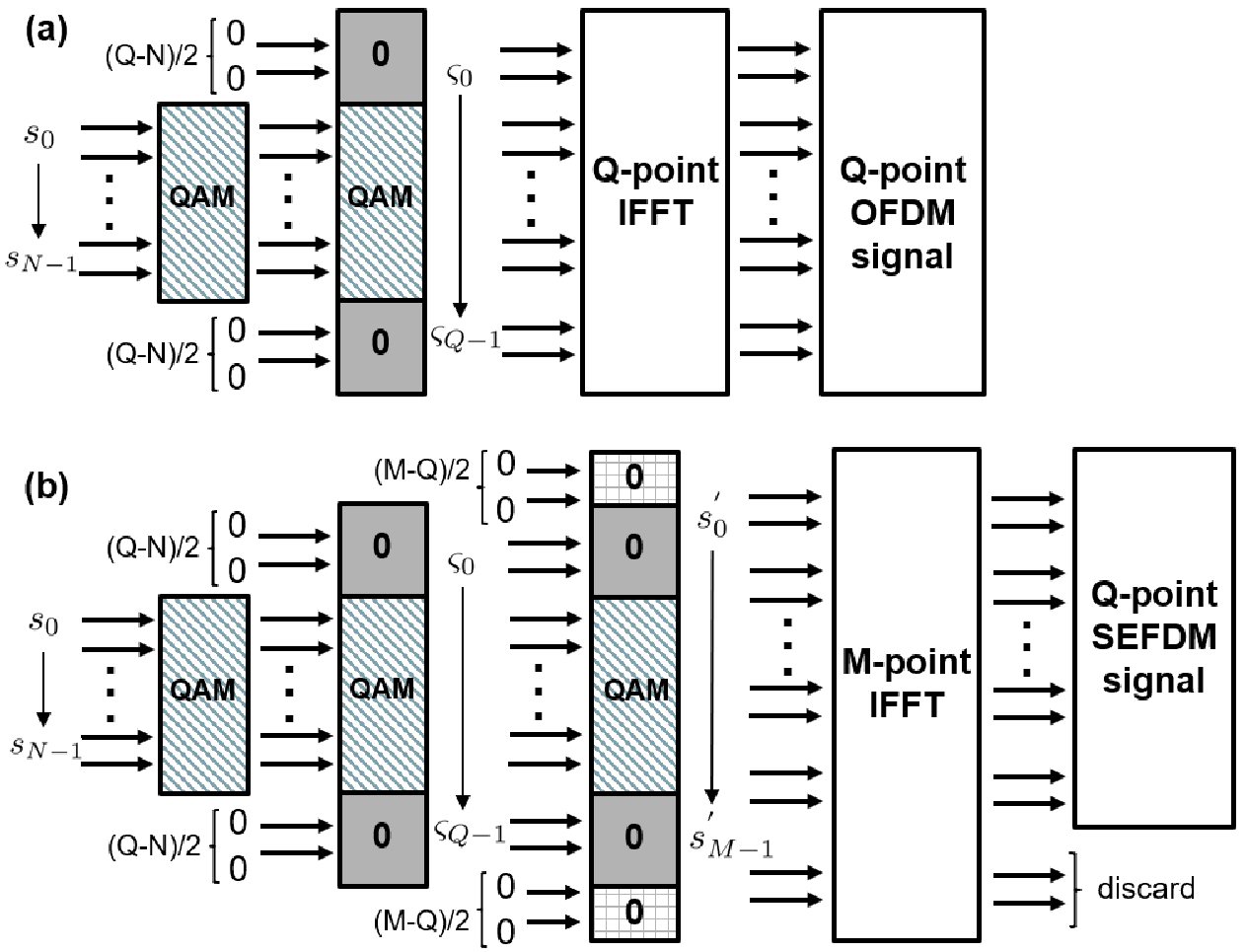}
\end{center}
\caption{Signal generation block diagram. (a) OFDM. (b) SEFDM.}
\label{Fig:WDS_TIFS_SEFDM_generation}
\end{figure}

The mathematical expression of an SEFDM signal is obtained by adding $\alpha$ in a typical OFDM signal as
\begin{equation}
\small
x_k=\frac{1}{\sqrt{Q}}\sum_{n=0}^{N-1}s_{n}\exp\left(\frac{j2{\pi}nk\alpha}{Q}\right),
\label{eq:SEFDM_discrete_signal}\end{equation}
where $\frac{1}{\sqrt{Q}}$ is the power scaling factor, $Q=\rho{N}$ is the number of time samples where $\rho$ is an oversampling factor and $N$ is the number of sub-carriers. $x_k$ is the $k^{th}$ time sample with the index $k=0,1,...,Q-1$. $s_n$ is the $n^{th}$ single-carrier symbol modulated on the $n^{th}$ sub-carrier.

Commonly, a signal requires protection guard bands on both sides. Therefore, in Fig. \ref{Fig:WDS_TIFS_SEFDM_generation}, the original input symbol vector $[s_0, s_1,...,s_{N-1}]$ is expanded to a Q-dimensional vector as
\begin{equation}
[\varsigma_0,\varsigma_1,...,\varsigma_{Q-1}]=[\underbrace{0,...,0}_{(Q-N)/2},s_0,s_1,...,s_{N-1},\underbrace{0,...,0}_{(Q-N)/2}].
\label{eq:OFDM_signal_padding}\end{equation}

Then a Q-point \ac{IFFT} is applied in Fig. \ref{Fig:WDS_TIFS_SEFDM_generation}(a) to modulate the vector $[\varsigma_0,\varsigma_1,...,\varsigma_{Q-1}]$ leading to a Q-point OFDM symbol. For SEFDM signal generation, equation \eqref{eq:SEFDM_discrete_signal} will be transformed into
\begin{equation}
\small
x_k=\frac{1}{\sqrt{Q}}\sum_{n=0}^{Q-1}\varsigma_{n}\exp\left(\frac{j2{\pi}nk\alpha}{Q}\right).
\label{eq:SEFDM_discrete_signal_expand}\end{equation}

It is clear that the direct operation of \eqref{eq:SEFDM_discrete_signal_expand} will result in high computational complexity due to the existence of $\alpha$. To remove the effect of $\alpha$ and directly use \ac{IFFT} for SEFDM signal generation, the vector $[\varsigma_0,\varsigma_1,...,\varsigma_{Q-1}]$ has to be further expanded to a longer vector as shown in Fig. \ref{Fig:WDS_TIFS_SEFDM_generation}(b) with the following operation
\begin{equation}
[s^{'}_0,s^{'}_1,...,s^{'}_{M-1}]=[\underbrace{0,...,0}_{(M-Q)/2},\varsigma_0,\varsigma_1,...,\varsigma_{Q-1},\underbrace{0,...,0}_{(M-Q)/2}],
\label{eq:SEFDM_signal_padding_2}\end{equation}
where a new parameter $M=Q/\alpha$ is defined. $M$ should be rounded to its closest integer. A vector of $(M-Q)/2$ zeros are padded on both sides of $\varsigma$. Therefore, the original signal generation expression will be transformed into an M-point IFFT operation demonstrated in Fig. \ref{Fig:WDS_TIFS_SEFDM_generation}(b) as
\begin{equation}
\small
x^{'}_k=\frac{1}{\sqrt{M}}\sum_{n=0}^{M-1}s^{'}_{n}\exp\left(\frac{j2{\pi}nk}M\right),\label{eq:signal_single_IFFT_pad_zeros}\end{equation}
where $n,k=[0,1,...,M-1]$. The output will be truncated with only $Q$ samples reserved while the rest of the samples are discarded.

To simplify the expression, a matrix format of the signal generation in \eqref{eq:SEFDM_discrete_signal} is defined as
\begin{equation}
\mathbf{x}=\mathbf{F}\mathbf{s},\label{eq:SEFDM_signal_matrix_format}\end{equation}
where $\mathbf{s}$ is the N-dimensional signal vector, $\mathbf{F}$ is the $Q\times{N}$ sub-carrier matrix, in which each element is represented by $\exp\left(\frac{j2{\pi}nk\alpha}{Q}\right)$.

{After going through a wireless channel denoted by a $Q\times{Q}$ channel matrix $\mathbf{H}$ and \ac{AWGN} $\mathbf{w}$, the received signal is expressed as
\begin{equation}
\mathbf{y}=\mathbf{H}\mathbf{x}+\mathbf{w}.\label{eq:simple_SEFDM_signal_H_AWGN}\end{equation}}

{It is noted that before any further signal processing after \eqref{eq:simple_SEFDM_signal_H_AWGN}, the channel effect of $\mathbf{H}$ has to be equalized by multiplying with the inverse of the channel matrix as
\begin{equation}\label{eq:simple_SEFDM_equalization}
\hat{\mathbf{y}}=\mathbf{H}^{-1}\mathbf{H}\mathbf{x}+\mathbf{H}^{-1}\mathbf{w}=\mathbf{x}+\mathbf{z}.
\end{equation}}

By multiplying the signal using the complex conjugate demodulation matrix $\mathbf{F}^*=\exp\left(\frac{-j2{\pi}nk\alpha}{Q}\right)$, the demodulated signal is expressed 
\begin{equation}
\mathbf{r}=\mathbf{F^{*}}\mathbf{x}+\mathbf{F^{*}}\mathbf{z}=\mathbf{F^{*}}\mathbf{F}\mathbf{s}+\mathbf{F^{*}}\mathbf{z}=\mathbf{C}\mathbf{s}+\mathbf{z}_{\mathbf{F^*}},
\label{eq:demodulation_FDM_signal_perfect}\end{equation} 
where $\mathbf{C}$ is the $N\times{N}$ correlation matrix, which includes the self-created ICI information as
\begin{equation} \label{eq:corr_perfect}
\begin{split}
&c_{m,n}=\\
&\frac{sinc[\pi\alpha(m-n)]}{sinc[\pi\alpha(m-n)/Q]}\times\exp\left(\frac{j{\pi}\alpha(Q-1)(m-n)}{Q}\right).
\end{split}
\end{equation}

When $m=n$, all the auto-correlation diagonal elements $c_{m,n}$ equal one. When $m{\neq}n$, all the cross-correlation non-diagonal elements are not zero indicating the self-created \ac{ICI}. It is apparent that the ICI term is related to the value of $\alpha$, which is the principle for the WDS communication security.

\subsection{Security Metric} \label{subsec:security_metric}

{The principle of this work is to design waveform patterns that can confuse eavesdroppers. Therefore, to evaluate the robustness, instead of using non-AI security metric SNR, we use AI security metric SCA to indicate the capability of eavesdroppers to correctly identify a signal.}
\begin{equation}
SCA=\frac{1}{\lambda}\sum_{\nu=1}^{\lambda}\frac{N_{C}(\nu)}{N_{T}(\nu)},
\label{eq:secrecy_metric_WDS}\end{equation}
{where the number of signal classes is defined by $\lambda$. The larger value of $\lambda$, the more difficult for an eavesdropper to successfully identify a signal pattern. To have solid evaluations, in each signal class with the index of $\nu$, a total number of $N_{T}$ symbols are tested. Among $N_{T}$ symbols, $N_{C}$ symbols can be correctly identified by an eavesdropper. The ratio of $N_{C}$ and $N_{T}$ indicates classification accuracy for one signal class. The final accuracy is obtained by averaging the results from $\lambda$ signal classes. A small value of $SCA$ indicates a low classification accuracy at Eve, which leads to the failure of signal detection and prevents accurate adversarial AI attacks.}

\subsection{Signal Classification Principle} \label{subsec:signal_classification_principle}

Signal classification is to identify different signal formats associated with the value of $\alpha$. A perfect signal classification will determine the accurate demodulation matrix $\mathbf{F}^*$ in \eqref{eq:demodulation_FDM_signal_perfect} and further determine the characteristics of $\mathbf{C}$. An imperfect signal classification will mistakenly use a wrong demodulation matrix as 
\begin{equation}
\tilde{\mathbf{r}}=\tilde{\mathbf{F}}^{*}\mathbf{x}+\tilde{\mathbf{F}}^{*}\mathbf{z}=\tilde{\mathbf{F}}^{*}\mathbf{F}\mathbf{s}+\tilde{\mathbf{F}}^{*}\mathbf{z}=\tilde{\mathbf{C}}\mathbf{s}+\mathbf{z}_{\tilde{\mathbf{F}}^*},
\label{eq:demodulation_FDM_signal}\end{equation} 
where $\tilde{\mathbf{F}}^{*}$ is the incorrect demodulation sub-carrier matrix caused by misclassification. Compared to the ideal matrix ${\mathbf{F}}^{*}$ in \eqref{eq:demodulation_FDM_signal_perfect}, a BCF offset $\Delta{\alpha}$ will exist in the imperfect $\tilde{\mathbf{F}}^{*}$ with the new expression as $\tilde{\mathbf{F}}^{*}=\exp\left(\frac{-j2{\pi}nk(\alpha+\Delta{\alpha})}{Q}\right)$. The mismatch between $\tilde{\mathbf{F}}^{*}$ and ${\mathbf{F}}$ will cause an imperfect estimate of $\tilde{\mathbf{C}}$ as
\begin{equation}\label{eq:corr_imperfect}
\small
\begin{split}
&\tilde{c}_{m,n}=\\
&\frac{sinc[\pi({\alpha_T}m-{\alpha_R}n)]}{sinc[\pi({\alpha_T}m-{\alpha_R}n)/Q]}\times\exp\left(\frac{j{\pi}(Q-1)({\alpha_T}m-{\alpha_R}n)}{Q}\right),
\end{split}
\end{equation}
where $\alpha_T$ is the BCF at the transmitter and $\alpha_R=\alpha_T+\Delta{\alpha}$ is the incorrect BCF at the receiver.

The traditional and optimal classification method is maximum likelihood, which has been investigated for modulation classification in \cite{classification_likelihood_AMC_TWC_2009, classification_likelihood_AMC_2011}. The likelihood function, with perfect knowledge of all parameters except modulation format, is expressed as  
\begin{equation}\small
L_f(r|\mathfrak{M},\sigma)=\frac{1}{P}\prod_{n=0}^{N-1}\sum_{p=0}^{P-1}\frac{1}{2\pi{\sigma^2}}\exp\left(-\frac{|r_n-\mathfrak{M}(i,p)|^2}{2\sigma^2}\right),
\label{eq:likelihood_function}\end{equation}
where $\mathfrak{M}$ represents the modulation class, $\mathfrak{M}(i,p)$ indicates the $p^{th}$ constellation symbol in the $i^{th}$ modulation scheme. There are $P$ constellation points for each modulation. $\sigma^2$ is noise variance when \ac{AWGN}  is considered and $r_n$ is the $n^{th}$ single-carrier complex symbol.

The optimal solution is obtained via maximizing the likelihood function \eqref{eq:likelihood_function} by attempting all the potential modulation candidates as
\begin{eqnarray}
\hat{\mathfrak{M}}=\arg\max_{\mathfrak{M}(i)\in \Theta}L_f(r|\mathfrak{M},\sigma),\label{eq:maximum-likelihood-decision}\end{eqnarray}
where $\Theta$ indicates all the potential candidates for the $i^{th}$ modulation format.

The traditional maximum likelihood method is not realistic for non-orthogonal signal classification. Therefore, intelligent classification using artificial intelligence would be a potential solution. Deep learning based \ac{CNN} has been investigated for single-carrier modulation classification in \cite{OShea_classification_2018} with competitive accuracy relative to the maximum likelihood method. The automatic learning CNN classifier has also been tested for non-orthogonal multi-carrier signal classification in \cite{tongyang_VTC2020_DL_classification}. Therefore, the CNN model will be used for eavesdropping signal classification in this work.

\begin{table}[t!]
\centering
\caption{ Hardware Complexity Analysis (uplink channel from Alice to Bob)}
\begin{tabular}{ | c | c | } \hline

$\mathbf{Framework}$ & Hardware \\  \hline \hline

Traditional PLS & RF Chain(multiple)     \\ 
(digital beamforming)  & Antenna(multiple)  \\ \hline

Traditional PLS &  RF Chain(multiple)      \\ 
(hybrid analog-digital beamforming)  & Antenna(multiple)  \\ \hline

Traditional PLS &  RF Chain(single)      \\ 
(analog beamforming)  & Antenna(multiple)  \\ \hline

WDS(Alice): &  RF Chain(single)      \\ 
User  & Antenna(single)  \\ \hline

WDS(Bob): &  RF Chain(single)      \\ 
Base Station  & Antenna(single)  \\ \hline

WDS(Eve): &  RF Chain(single/multiple)      \\ 
Eavesdropper  & Antenna(single/multiple)  \\ \hline

\end{tabular}
\label{tab:WDS_power_complexity_analysis}
\end{table}

\begin{figure}[t!]
\begin{center}
\includegraphics[scale=0.5]{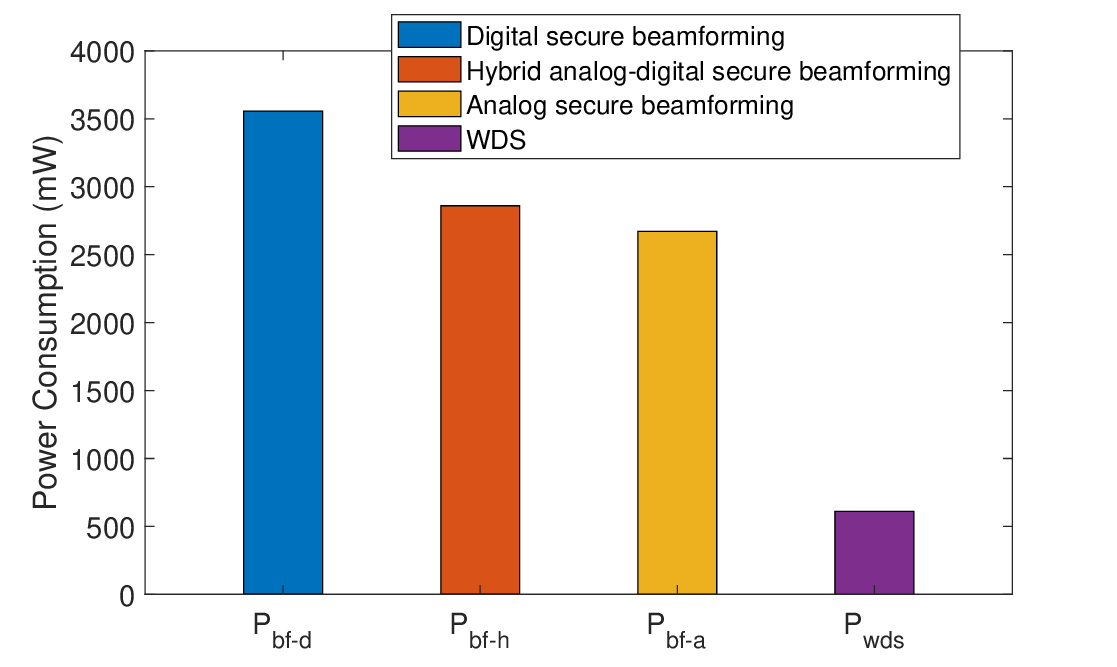}
\end{center}
\caption{Power consumption comparison for full-digital secure beamforming, hybrid analog-digital secure beamforming, analog secure beamforming and WDS frameworks.}
\label{Fig:MAMB_power_consumption}
\end{figure}

\subsection{Power Consumption Comparison} \label{sec:complexity_power_consumption}

The WDS framework aims for low-cost and resource-constrained communications where IoT is a matched application scenario. Most IoT traffic occurs at uplink channels where each IoT unit sends information back to base stations. {Therefore, we consider power consumption for uplink channel communications with the complexity comparison in Table \ref{tab:WDS_power_complexity_analysis} where hardware utilization for each scenario is compared. In the column of `Hardware', detailed hardware utilization is presented. In the bracket, `single' indicates one such component is needed while `multiple' indicates several such components have to be used. There is no specific data associated with Table \ref{tab:WDS_power_complexity_analysis} where this table only shows general hardware utilization for different scenarios.} The proposed multi-band WDS framework utilizes single-RF chain while traditional PLS has to employ multiple RF chains for digital beamforming, where the traditional solution consume more power \cite{GreenCom_Nature_Electronics_2017, GreenCom_6G_summit_2020}.

Based on the studies in \cite{PLS_power_consumption_TCOM_2021, PLS_power_consumption_JSTSP_2018}, the power consumption for digital beamforming $P_{bf-d}$, hybrid analog-digital beamforming $P_{bf-h}$, analog beamforming $P_{bf-a}$ and WDS $P_{wds}$ could be computed in the following
\begin{equation}\small
P_{bf-d}=P_{lo}+N_{rf-f}(P_{dac}+P_{mixer}+P_f+\frac{P_t}{\xi}),
\label{eq:power_beamforming_full}\end{equation}
\begin{equation}\small
P_{bf-h}=P_{lo}+N_{rf-h}(P_{dac}+P_{mixer}+P_f)+N_{ps}(P_{ps}+\frac{P_t}{\xi}),
\label{eq:power_beamforming_hybrid}\end{equation}
\begin{equation}\small
P_{bf-a}=P_{lo}+P_{dac}+P_{mixer}+P_f+N_{ps}(P_{ps}+\frac{P_t}{\xi}),
\label{eq:power_beamforming_analog}\end{equation}
\begin{equation}\small
P_{wds}=P_{lo}+P_{dac}+P_{mixer}+P_f+\frac{P_t}{\xi},
\label{eq:power_wds}\end{equation}
where $P_{lo}$, $P_{dac}$, $P_{mixer}$, $P_f$, $P_t$ and $P_{ps}$ indicate the power consumption for the local oscillator, \ac{DAC}, mixer, filter, transmit signal and phase shifter. $N_{rf-d}$ is the number of RF chains for digital beamforming, $N_{rf-h}$ is the number of RF chains for hybrid beamforming and $N_{ps}$ is the number of phase shifters. $\xi$ indicates the efficiency of a power amplifier. {Based on \cite{PLS_power_consumption_TCOM_2021, PLS_power_consumption_JSTSP_2018}, we set $P_{lo}$=22 mW, $P_{dac}$=170 mW, $P_{mixer}$=5 mW, $P_f$=14 mW, $P_t$=200 mW, $P_{ps}$=10 mW, $\xi$=50\%. Based on \cite{Tongyang_MU_MIMO_NB_IoT_2018, Tongyang_hybrid_precoding_2020}, we set $N_{rf-f}$=6, $N_{rf-h}$=2, $N_{ps}$=6.} The power consumption for each system design is compared in Fig. \ref{Fig:MAMB_power_consumption}, in which our proposed WDS framework can reduce power consumption by up to six times compared to traditional multi-antenna based secure beamforming techniques.

\section{Signal Detection} \label{sec:signal_detection}

In the framework in Fig. \ref{Fig:MAMB_eavesdropping_model}, the legitimate user Bob has correct signal detection because the signal pattern information is pre-known between Alice and Bob. However, signal detection at Eve would fail due to signal misclassification.

\subsection{WDS Signal Detection} \label{subsec:principle_signal_detection}

{Once the correlation matrix $\mathbf{C}$ is determined via either paired-key information} at Bob or signal classification at Eve, signal detection has to be operated to recover original signals from ICI. The optimal signal detection method is \ac{ML} while its computational complexity is exponentially increased when the number of sub-carriers increases. Its simplified version is \ac{SD} \cite{SD_orginal}, which searches for the optimal solution within a pre-defined space. 

The SD search for the optimal estimate $\mathbf{s}_{_{SD}}$ is defined as
\begin{eqnarray}\mathbf{s}_{_{SD}}=\arg\min_{{\mathbf{s}}\in O^N}\left\Vert {\mathbf{r}}-\mathbf{C}{{\mathbf{s}}}\right\Vert ^{2}\label{eq:sd}\le{g},\end{eqnarray}
where $O$ is the constellation cardinality and $O^{N}$ covers all possible solutions. $g$ is the pre-defined search radius and it equals the distance between the demodulated $\mathbf{r}$ and the hard-decision $\mathbf{s}_{_{ZF}}$. It is noted that the hard-decision $\mathbf{s}_{_{ZF}}$ is computed based on the \ac{ZF} criterion using a rounding function $\lfloor{.}\rceil$ as $\mathbf{s}_{_{ZF}}=\lfloor{\mathbf{C^{-1}}\mathbf{r}}\rceil$. Therefore, the search radius is defined as
\begin{eqnarray}{g}=\left\Vert \mathbf{r}-\mathbf{C}{\mathbf{s}_{_{ZF}}}\right\Vert ^{2}.\label{eq:SD_radius}\end{eqnarray}

The norm calculation in \eqref{eq:sd} can be re-formatted in \eqref{eq:SD_expanding} by substituting $\mathbf{p}=\mathbf{C^{-1}}\mathbf{r}$ where $\mathbf{p}$ is the soft-decision estimate of $\mathbf{s}$.
\begin{eqnarray}{\mathbf{s}}_{_{SD}}=\arg\min_{{{\mathbf{s}}}\in O^N}\{(\mathbf{p}-\mathbf{s})^{*}\mathbf{C}^{*}\mathbf{C}(\mathbf{p}-\mathbf{s})\}\le{{g}}.\label{eq:SD_expanding}\end{eqnarray} 

The expression can be further simplified using Cholesky decomposition \cite{Cholesky_Decomposition_1985} via $chol\{\mathbf{C}^{*}\mathbf{C}\}=\mathbf{L}^{*}\mathbf{L}$, where $\mathbf{L}$ is an $N\times N$ upper triangular matrix. Therefore, \eqref{eq:SD_expanding} can be re-written as
\begin{eqnarray}{\mathbf{s}}_{_{SD}}=\arg\min_{{{\mathbf{s}}}\in O^N}\left\Vert \mathbf{L}(\mathbf{p}-\mathbf{s})\right\Vert ^{2}\le{{g}}.\label{eq:SD_chol}\end{eqnarray}

The triangular structure of $\mathbf{L}$ can simplify \eqref{eq:SD_chol} into $N$ steps with the following expression
\begin{eqnarray}
\begin{split}
g{\geq}&(l_{_{N-1,N-1}}(p_{_{N-1}}-s_{_{N-1}}))^2+(l_{_{N-2,N-2}}(p_{_{N-2}}-\\
&s_{_{N-2}})+l_{_{N-2,N-1}}(p_{_{N-1}}-s_{_{N-1}}))^2+...,
\label{eq:SD_radius_Cholesky_steps}
\end{split}
\end{eqnarray}
where $l_{i,j}$, $p_i$ and $s_i$ are the elements of $\mathbf{L}$, $\mathbf{p}$ and $\mathbf{s}$ in \eqref{eq:SD_chol}, respectively. To study each term in \eqref{eq:SD_radius_Cholesky_steps}, the N-dimensional expression is divided into $N$ independent one-dimensional terms. The $(N-1)^{th}$ inequality term is thus represented as
\begin{eqnarray}l_{_{N-1,N-1}}^2({p}_{_{N-1}}-\mathbf{s}_{_{N-1}})^2\leq{g_{_{N-1}}=g}.\label{eq:SD_radius_Cholesky_N}\end{eqnarray}

Therefore, the search range for the $(N-1)^{th}$ dimension is derived as
\begin{eqnarray}
\lceil{-\frac{\sqrt{g_{_{N-1}}}}{l_{_{N-1,N-1}}}+{p}_{_{N-1}}}\rceil\leq{s_{_{N-1}}}\leq\lfloor{\frac{\sqrt{g_{_{N-1}}}}{l_{_{N-1,N-1}}}+{p}_{_{N-1}}}\rfloor,
\label{eq:SD_radius_Cholesky_N_LB_UB}
\end{eqnarray}
where $\lceil\cdotp\rceil$ $\lfloor\cdotp\rfloor$ denote rounding operations to the nearest larger and smaller integers, respectively. 

Therefore, the left term of \eqref{eq:SD_radius_Cholesky_N_LB_UB} indicates a hard lower bound (H-LB) while the right term indicates a hard upper bound (H-UB). It is clearly seen that an accurate estimate of $s_{_{N-1}}$ is related to $g_{_{N-1}}$, $l_{_{N-1,N-1}}$ and $p_{_{N-1}}$, which are all determined by the accurate estimate of $\mathbf{C}$.

After the search at the $(N-1)^{th}$ dimension, the search radius $g_{_{N-2}}$ for the next dimension is updated as 
\begin{eqnarray}g_{_{N-2}}=g_{_{N-1}}-l_{_{N-1,N-1}}^2(p_{_{N-1}}-s_{_{N-1}})^2.\label{eq:SD_radius_Cholesky_N-1_radius}\end{eqnarray}

The search principle in \eqref{eq:SD_radius_Cholesky_N_LB_UB} and the radius update in \eqref{eq:SD_radius_Cholesky_N-1_radius} will be repeated until the last dimension. The final solution ${\mathbf{s}}_{_{SD}}$ is obtained as an N-dimensional vector that meets the condition in \eqref{eq:sd}. Each element estimation in ${\mathbf{s}}_{_{SD}}$ is dependent on the elements from its previous dimensions. The perfect knowledge of $\mathbf{C}$ plays an important role since an imperfect estimate of $\mathbf{C}$ will give a wrong decision interval in \eqref{eq:SD_radius_Cholesky_N_LB_UB} and might cause no solution at the end. Therefore, the first step signal classification is crucial to an eavesdropper who aims to decode signals.

\subsection{Impact of Imperfect Classification}

An imperfect signal classification will mislead the estimate of $\mathbf{C}$, which further gives inaccurate calculation of $\mathbf{L}$ in Cholesky decomposition. Therefore, the element $l_{i,j}$ in $\mathbf{L}$ will become $l_{i,j}+{\Delta}l$, where ${\Delta}l$ is the offset caused by imperfect signal classification. Meanwhile, since the soft-decision estimation follows $\mathbf{p}=\mathbf{C^{-1}}\mathbf{r}$, the new estimate of each element will become $p_i+{\Delta}p$ where ${\Delta}p$ is the offset caused by imperfect signal classification. It should be noted that signal misclassification will cause inaccurate $\hat{\mathbf{s}}_{_{ZF}}=\lfloor{(\mathbf{C}+\Delta{\mathbf{C}})^{-1}\mathbf{r}}\rceil$ as well. Therefore, the search space $g_i$ in \eqref{eq:SD_radius} will become $g_i+{\Delta}g$ where ${\Delta}g$ is the offset caused by imperfect signal classification.

The above imperfect estimates will jointly cause inaccurate estimate of $\mathbf{s}$. The lower bound and upper bound in \eqref{eq:SD_radius_Cholesky_N_LB_UB} will be improperly biased to 
\begin{eqnarray}\small
LB=\lceil{-\frac{\sqrt{g_{_{N-1}}+{\Delta}g}}{l_{_{N-1,N-1}}+{\Delta}l}+p_{_{N-1}}+{\Delta}p}\rceil. 
\label{eq:SD_radius_Cholesky_N_LB_imperfect}
\end{eqnarray}
\begin{eqnarray}\small
UB=\lfloor{\frac{\sqrt{g_{_{N-1}}+{\Delta}g}}{l_{_{N-1,N-1}}+{\Delta}l}+p_{_{N-1}}+{\Delta}p}\rfloor.
\label{eq:SD_radius_Cholesky_N_UB_imperfect}
\end{eqnarray}

Therefore, the variations of ${\Delta}g$, ${\Delta}l$ and ${\Delta}p$, due to imperfect signal classification, will cause signal detection failure.

\section{Secure Multi-Band Framework} \label{sec:multi-band_WDS_framework}

\begin{figure}[t!]
\begin{center}
\includegraphics[scale=0.31]{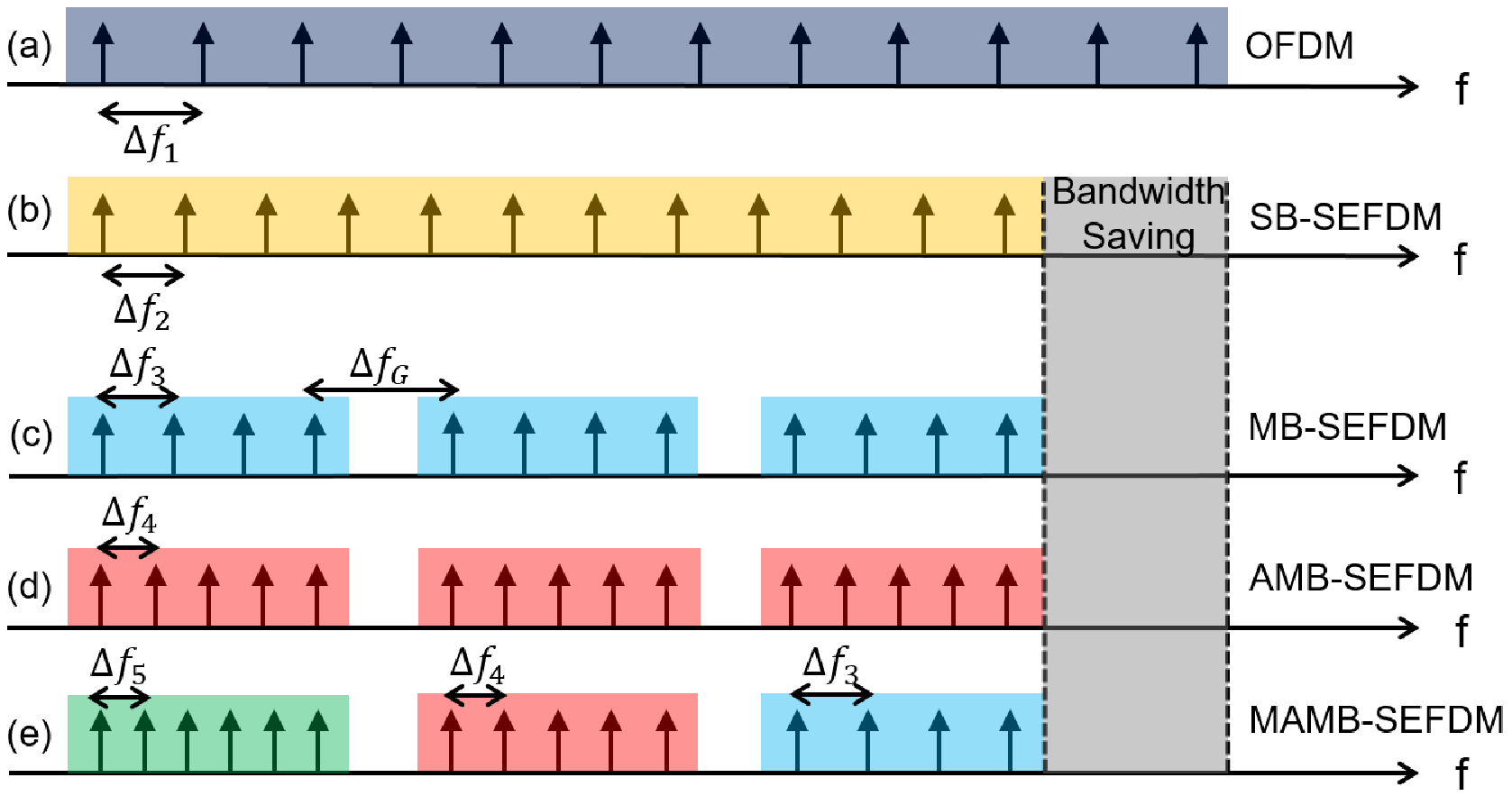}
\end{center}
\caption{Spectral illustration for (a) OFDM, (b) SB-SEFDM, (c) MB-SEFDM, (d) AMB-SEFDM, (e) MAMB-SEFDM. Each impulse in each sub-figure indicates one sub-carrier and each coloured rectangular block indicates a signal band or a sub-band.}
\label{Fig:MAMB_waveform_arch_all}
\end{figure}

To ensure a joint secure and detectable communication system, the signal waveform has to be modified. This section will investigate four WDS signal waveform architectures in Fig. \ref{Fig:MAMB_waveform_arch_all}, namely \ac{SB-SEFDM}, \ac{MB-SEFDM}, \ac{AMB-SEFDM} and \ac{MAMB-SEFDM}.

\subsection{Single-Band} \label{subsec:singleband}

The WDS framework was initially designed for single-band signals. In this case, sub-carriers are packed consecutively without empty guard bands. The traditional \ac{SB-SEFDM} signal architecture is presented in Fig. \ref{Fig:MAMB_waveform_arch_all}(b). To simplify the illustration, each impulse represents a sub-carrier. For a better demonstration, only a partial number of sub-carriers and sub-bands are presented. It should be noted that all the designs in Fig. \ref{Fig:MAMB_waveform_arch_all} have the same sub-carrier bandwidth. The only difference is the sub-carrier spacing. In order to achieve bandwidth compression, the sub-carrier spacing for SB-SEFDM should satisfy $\Delta{f_2}<\Delta{f_1}$, where $\Delta{f_1}$ and $\Delta{f_2}$ indicate the sub-carrier spacing of OFDM and SB-SEFDM, respectively.

According to 4G \cite{3GPPrelease14} and 5G \cite{3GPPrelease15} standards, a multi-carrier signal bandwidth is defined by the multiplication of sub-carrier spacing $\Delta{f}$ and the number of sub-carriers $N$. Therefore, the spectral bandwidths for the cases in Fig. \ref{Fig:MAMB_waveform_arch_all}(a) and Fig. \ref{Fig:MAMB_waveform_arch_all}(b) are defined by $B_{_{OFDM}}=N\Delta{f}$ and $B_{_{SB-SEFDM}}=\alpha{N\Delta{f}}$ respectively.

The single-band SB-SEFDM signal architecture might challenges signal detection since the sophisticated SD detector has to be applied resulting in exponentially increased computational complexity especially when the size of a signal is scaled up. Thus, communication security is ensured such that eavesdroppers cannot decode signals easily but at the cost of complicating legitimate user signal recovery as well.

\subsection{Multi-Band} \label{subsec:multiband}

The principle of the multi-band signal architecture, shown in Fig. \ref{Fig:MAMB_waveform_arch_all}(c), is to partition the single-band signal into multiple sub-bands with an empty sub-carrier between two adjacent sub-bands. The purpose of the protection gap is to mitigate inter-band interference. In this case, each sub-band signal can be recovered separately using the SD detector leading to reduced computational complexity.  

The total occupied spectral bandwidth of the multi-band signal is equivalent to that of a typical single-band signal. Due to one empty sub-carrier as the protection gap $\Delta{f_G}=2\Delta{f_3}$ between two adjacent sub-bands in Fig. \ref{Fig:MAMB_waveform_arch_all}(c), the sub-carrier spacing in each sub-band has to be further squeezed leading to the spacing $\Delta{f_3}<\Delta{f_2}<\Delta{f_1}$. The effective spectral bandwidth of MB-SEFDM is defined as
\begin{equation} \label{eq:multi-SEFDM_bandwidth}
\small
B_{_{MB-SEFDM}}=\beta{(N+\frac{N}{N_B}-1)\Delta{f}},
\end{equation}
where $N_B$ is the number of sub-carriers in each sub-band and $\beta$ indicates the sub-band bandwidth compression factor. To ensure the same occupied spectral bandwidth $B_{_{MB-SEFDM}}=B_{_{SB-SEFDM}}$, the sub-band $\beta$ is calculated as
\begin{equation}\beta=\frac{\alpha{N}}{N+\frac{N}{N_B}-1}.\label{eq:sub-band_BCF}\end{equation}

The mathematical expression of the multi-band SEFDM signal is given by
\begin{equation}\label{eq:multi_SEFDM_signal}
\small
\begin{split}
&x_k=\\
&\frac{1}{\sqrt{Q}}\sum_{l_{B}=0}^{{\frac{N}{N_B}}-1}\sum_{i=0}^{N_B-1}s_{_{i+{l_{B}}N_B}}\exp\left(\frac{j2{\pi}k\beta(i+{l_{B}}(N_B+1))}{Q}\right),
\end{split}
\end{equation}
where $s_{_{i+{l_{B}}N_B}}$ is the $i^{th}$ single-carrier symbol modulated in the ${l_{B}}^{th}$ sub-band.

To directly use \ac{IFFT} for the multi-band SEFDM signal generation, the raw symbol mathematical expression in \eqref{eq:multi_SEFDM_signal} has to be updated to
\begin{equation}
\small
{s^{''}_m = s^{''}_{n+\lfloor{\frac{n}{N_B}}\rfloor} = \left\{
  \begin{array}{l l}
    s_n & \quad \text{$0\ {\leq}\ n<N$} \\
    0 & \quad \text{$otherwise$}
  \end{array} \right.},
\label{eq:multi_SEFDM_signal_regeneration}\end{equation}
where related parameters are defined below
\begin{equation}
\small
{\left\{
  \begin{array}{l l}
    n & \quad \text{$=i+{l_{B}}N_B$} \\
    m & \quad \text{$=i+{l_{B}}(N_B+1)=n+l_{B}$} \\
    l_{B} & \quad \text{$=\lfloor{\frac{n}{N_B}}\rfloor$}
  \end{array} \right.}.
\label{eq:multi_SEFDM_signal_regeneration_pre_define}\end{equation}

Therefore, the original multi-band signal expression in \eqref{eq:multi_SEFDM_signal} is converted to a new expression as
\begin{equation}\label{eq:multi_SEFDM_ifft_discrete_signal}
\small
x_k=\frac{1}{\sqrt{Q}}\sum_{m=0}^{N+\frac{N}{N_B}-2}s^{''}_{m}\exp\left(\frac{j2{\pi}mk\beta}{Q}\right).
\end{equation}

Following the same zero padding method in \eqref{eq:SEFDM_signal_padding_2}, a new input symbol vector is generated as
\begin{equation}
\small
s^{'''}_{m} = \left\{
  \begin{array}{l l}
    0 & \quad \text{$0{\leq}m<(M-Q^{'})/2$}\\
    s^{''}_{m} & \quad \text{$(M-Q^{'})/2{\leq}m<(M+Q^{'})/2$}\\
    0 & \quad \text{$(M+Q^{'})/2{\leq}m<M$}
  \end{array} \right.,
\label{eq:symbol_vector_single_IFFT_SEFDM_multiband}\end{equation} 
where $Q^{'}=N+\frac{N}{N_B}-1$, $M={Q}/\beta$ is rounded to its closest integer. The expression in \eqref{eq:multi_SEFDM_ifft_discrete_signal} is therefore adjusted to a new form as
\begin{equation}
\small
x^{''}_k=\frac{1}{\sqrt{M}}\sum_{m=0}^{M-1}s^{'''}_{m}\exp\left(\frac{j2{\pi}mk}{M}\right),\label{eq:multiband_signal_single_IFFT_pad_zeros}\end{equation}
where $m,k=[0,1,...,M-1]$. The output is truncated with only $Q$ samples reserved while the rest of the samples are discarded.

\subsection{Adaptive Multi-Band}

The multi-band signal architecture simplifies signal detection. However, the challenge of the multi-band signal architecture is that eavesdroppers can filter and extract each sub-band and operate signal classification for each one. To enhance multi-band communication security, an adaptive multi-band signal architecture is proposed in Fig. \ref{Fig:MAMB_waveform_arch_all}(d).

Modifying spectral features of a signal would effectively prevent unauthorized signal feature learning and format identification. It is observed from Fig. \ref{Fig:MAMB_waveform_arch_all}(d) that the overall occupied spectral bandwidth is similar to the traditional MB-SEFDM but with further reduced bandwidth compression factor leading to $\Delta{f_4}<\Delta{f_3}<\Delta{f_2}<\Delta{f_1}$. The scheme in Fig. \ref{Fig:MAMB_waveform_arch_all}(d) would mislead eavesdroppers to classify an AMB signal of $\beta_0$ into an MB signal of $\beta_1$ due to their similar spectral characteristics. Meanwhile, the AMB signal architecture in Fig. \ref{Fig:MAMB_waveform_arch_all}(d) achieves a higher data rate than the MB signal in Fig. \ref{Fig:MAMB_waveform_arch_all}(c).

Considering an example comparison including three types of signals where the bandwidth compression factors for the signals in each sub-band satisfy $\beta_{2}<\beta_{1}<\beta_{0}$. To make three signals similar, more sub-carriers will be packed in $\beta_{1}$, $\beta_{2}$ relative to $\beta_{0}$. The sub-carrier packing strategy is
\begin{equation}\label{eq:AMB_subcarrier_packing}
\small
B_{_{sub}}={\beta_0}N_B\Delta{f}={\beta_1}(N_B+\Delta{N_1})\Delta{f}={\beta_2}(N_B+\Delta{N_2})\Delta{f},\end{equation}
where $B_{_{sub}}$ is the bandwidth for one sub-band, $\Delta{N_1}$ is the number of additional sub-carriers per sub-band that have to be packed for $\beta_{1}$ relative to $\beta_{0}$ and $\Delta{N_2}$ is the number of additional sub-carriers per sub-band that have to be packed for $\beta_{2}$ relative to $\beta_{0}$. In this case, the spectral bandwidth per sub-band for the three SEFDM signals would be similar and can easily cause eavesdropping misclassification.

Due to additional sub-carrier packing, the original multi-band signal in \eqref{eq:multi_SEFDM_signal} is modified to a new format as
\begin{equation}\label{eq:aligned_multi_SEFDM_signal}
\small
\begin{split}
x_k=\frac{1}{\sqrt{Q}}&\sum_{l_{B}=0}^{{\frac{N+\Delta{N}}{N_B+\Delta{N_B}}}-1}\sum_{i=0}^{N_B+\Delta{N_B}-1}s_{_{i+{l_{B}}(N_B+\Delta{N_B})}}\\
&\times\exp\left(\frac{j2{\pi}k\beta(i+{l_{B}}(N_B+\Delta{N_B}+1))}{Q}\right),
\end{split}
\end{equation}
where the number of sub-carriers in each sub-band is increased to $N_B+\Delta{N_B}$ and the total number of sub-carriers is increased to $N+\Delta{N}$. However, the number of sub-bands maintains the same with the following relationship
\begin{equation}\label{eq:AMB_number_sub_band}
\small
\frac{N+\Delta{N}}{N_B+\Delta{N_B}}=\frac{N}{N_B}.
\end{equation}

Signal generation for the AMB signal in \eqref{eq:aligned_multi_SEFDM_signal} is straightforward via IFFT following the similar operations from \eqref{eq:multi_SEFDM_signal_regeneration} to \eqref{eq:multiband_signal_single_IFFT_pad_zeros} except that more data sub-carriers are required by \eqref{eq:aligned_multi_SEFDM_signal}.

\subsection{Mixed Adaptive Multi-Band}

To enhance further communication security, a \ac{MAMB} signal waveform design is considered to flexibly tune BCF in each sub-band, where each sub-band has different BCF configurations but the overall effective BCF maintains the same. In Fig. \ref{Fig:MAMB_waveform_arch_all}(e), each independent sub-band has different number of sub-carriers, by adjusting sub-carrier spacing, the spectral bandwidth for each sub-band and the total occupied spectral bandwidth maintain the same leading to more confusions to eavesdroppers.

To confuse eavesdroppers, the sub-band BCF can be intentionally tuned with various patterns. Since each sub-band has a unique BCF, signal generation using a single-IFFT might be unrealistic. Therefore, multiple IFFTs have to be used and the number of IFFTs depends on the number of sub-bands. The composite MAMB signal, including all sub-bands, is represented as the following
\begin{equation}\label{eq:adaptive_aligned_multi_SEFDM_signal}
\small
\begin{split}
x_k=&\frac{1}{\sqrt{Q}}\sum_{i=0}^{N_B+\Delta{N_0}-1}s_{0i}\times\exp\left(\frac{j2{\pi}k\beta_0i}{Q}\right)\\
&+\frac{1}{\sqrt{Q}}\sum_{i=0}^{N_B+\Delta{N_1}-1}s_{1i}\times\exp\left(\frac{j2{\pi}k\beta_1(i+\varepsilon_0))}{Q}\right)\\
&+...\\
&+\frac{1}{\sqrt{Q}}\sum_{i=0}^{N_B+\Delta{N_{\Theta}}-1}s_{{\Theta}i}\times\exp\left(\frac{j2{\pi}k\beta_{\Theta}(i+\varepsilon_{\Theta}))}{Q}\right),
\end{split}
\end{equation}
where $\Theta=N/N_B-1$ is the maximum number of sub-band index, $s_{0i}$ indicates the $i^{th}$ symbol in the first sub-band and $s_{{\Theta}i}$ indicates the $i^{th}$ symbol in the $\Theta^{th}$ sub-band. $\beta_0$ is the sub-band BCF in the first sub-band and $\beta_\Theta$ is the sub-band BCF in the $\Theta^{th}$ sub-band. 

Since each sub-band has to be perfectly aligned without causing any spectral feature difference, each sub-band has to uniquely pack extra sub-carriers (i.e. $\Delta{N_0},\Delta{N_1},...,\Delta{N_{\Theta}}$) and has to be adaptively offset in frequency domain (i.e. $\varepsilon_0,\varepsilon_1,...,\varepsilon_{\Theta}$). It should be noted that the frequency offset for each sub-band alignment can be easily implemented by adaptively adding zeros to input symbol vectors (i.e. $s_{0i},s_{1i},...,s_{{\Theta}i}$) similar to the operations in \eqref{eq:OFDM_signal_padding} and \eqref{eq:SEFDM_signal_padding_2}. Then similar operations will be followed from \eqref{eq:multi_SEFDM_signal_regeneration} to \eqref{eq:multiband_signal_single_IFFT_pad_zeros} before the direct use of IFFT for signal generation.

\section{Pattern Key Generation} \label{sec:signal_pattern_exchange}

To ensure the communication reliability between legitimate users, the signal pattern key, has to be known between Alice and Bob. {However, it is impractical to exchange a large number of pattern information between legitimate users in each communication session. Therefore, an efficient way to generate pattern keys at both sides is of great importance.}

{A paired-key generator is proposed in this work.} The idea is to design a signal pattern generator that will be deployed at both the transmitter and the receiver. A chaotic dynamic system \cite{Chaos_nature_1976} can generate a random-like but reproducible chaotic sequence, which will be a simple solution for the WDS signal pattern generator. A discrete-time dynamical system is defined with the following state equation
\begin{equation}
\phi_{k+1}=f(\phi_k),\label{eq:chaotic_equ}\end{equation}
where $0<\phi_k<1$ indicates the value at the $k^{th}$ state and $0<\phi_{k+1}<1$ indicates the value at the $(k+1)^{th}$ state. $f(\cdotp)$ represents a chaotic map, which is used to produce sequence bits at different states. It is noted that the value of next state is highly dependent on its previous state. There are various chaotic maps and the commonly used one is logistic map, which is defined in \cite{Chaos_nature_1976} as
\begin{equation}
\phi_{k+1}=\gamma{\cdotp}\phi_k(1-\phi_k),\label{eq:chaotic_logistic_map}\end{equation}
{where $\gamma$ is the bifurcation parameter with values $1<\gamma<4$ defined by \cite{Chaois_comm_TCOM_1994}. The value of $\gamma$ determines the feature of a generated sequence. With a larger value of $\gamma$, the generated sequence is non-periodic and non-converging.} The studies in \cite{Chaois_comm_TCOM_1994} have proved that a minor change of the three factors will produce a completely different sequence.

\begin{figure}[t]
\begin{center}
\includegraphics[scale=0.51]{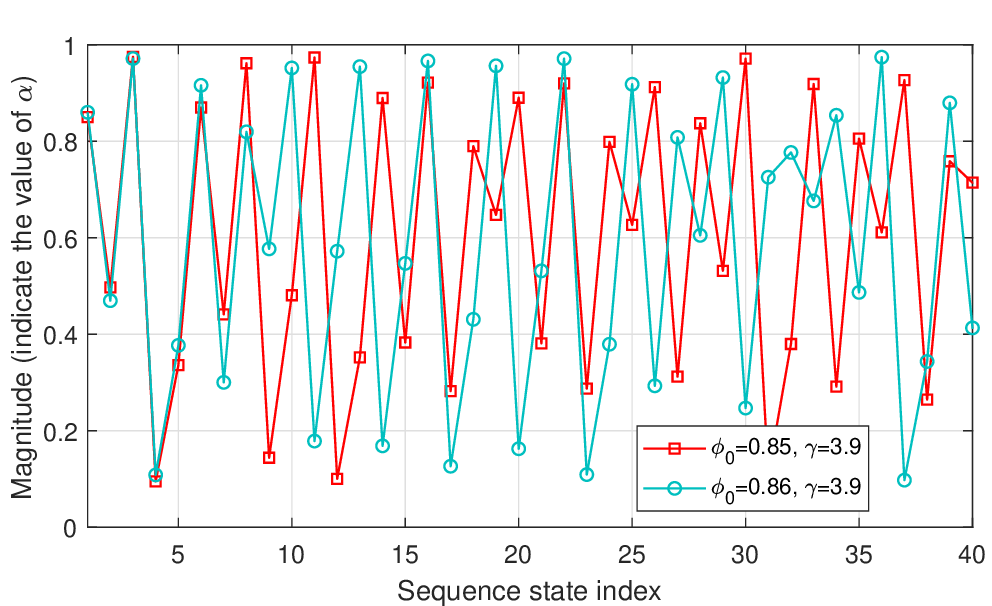}
\end{center}
\caption{Chaotic sequence illustration for two configurations with the minor difference in initial state $\phi_0$.}
\label{Fig:Chaos_seq_illustration_comparison}
\end{figure}

To show the pattern sequence generation mechanism, we compare two systems in Fig. \ref{Fig:Chaos_seq_illustration_comparison}. The first one is configured with $\gamma$=3.9, initial state $\phi_0$=0.85 and the logistic map following \eqref{eq:chaotic_logistic_map}. The second system has the same configurations except that the initial state is slightly increased to $\phi_0$=0.86. With such a minor difference, two systems will produce different sequences in Fig. \ref{Fig:Chaos_seq_illustration_comparison}, which can effectively prevent eavesdroppers from using exhaustive methods to guess the sequence. 

{The random-like sequence, as shown in Fig. \ref{Fig:Chaos_seq_illustration_comparison}, helps to generate pattern key $\alpha$. To implement the generation algorithm, a threshold $0<\eta<1$ is introduced to decide which pattern key should be generated. For example, to generate a pattern key sequence with $\alpha$=(0.9, 0.85, 0.8) using Fig. \ref{Fig:Chaos_seq_illustration_comparison}, a threshold $\eta$=0.75 could be used where only the values beyond the threshold $\eta$ is considered. For any values between 0.75 and 0.8, the key is $\alpha$=0.8; for any values between 0.8 and 0.85, the key is $\alpha$=0.85; for any values between 0.85 and 0.9, the key is $\alpha$=0.9. In this case, a pattern key sequence including $\alpha$=(0.9, 0.85, 0.8), is obtained.} 

The cooperation of the bifurcation parameter $\gamma$, chaotic map $f(\cdotp)$, initial state $\phi_0$ and pattern threshold $\eta$ will enable an efficient and secure pattern index generation scheme. An eavesdropper will not easily obtain an accurate pattern index sequence since a minor change of each parameter will produce a completely different sequence.

\section{Classifier Training and System Performance} \label{sec:simulation_results}

\subsection{Classifier Training}

\begin{figure}[t!]
\begin{center}
\includegraphics[scale=0.24]{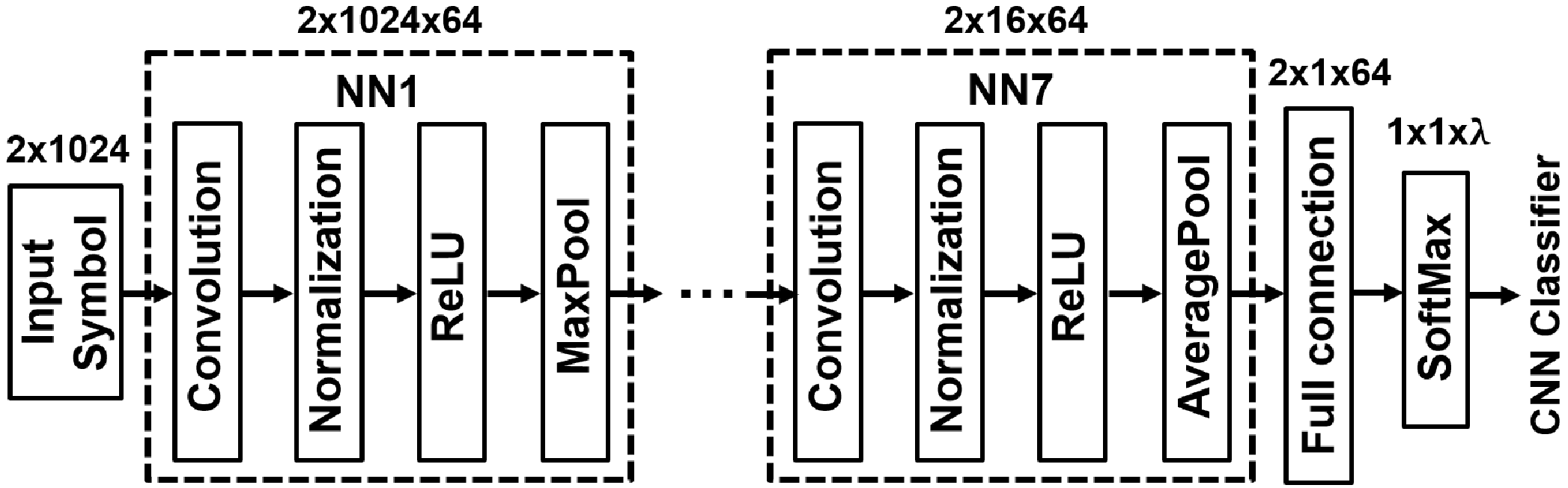}
\end{center}
\caption{CNN signal classifier architecture.}
\label{Fig:WDS_multiband_CNN_architecture}
\end{figure}

{The trained CNN architecture is presented in Fig. \ref{Fig:WDS_multiband_CNN_architecture} where seven convolutional layers are stacked for automatic feature extraction. The dimension of each layer is presented above each neural network sub-block. Each training symbol is configured to have 2048 complex time samples. To have a robust classifier, 1024 training samples is randomly captured out of the 2048 time samples. Therefore, the input training symbol size is 2${\times}$1024 since a complex symbol has real and imaginary parts. To avoid overfitting in the neural network training, a 50\% dropout ratio is configured.} To have a universal classifier that can generally identify signals at different noise conditions, the training signals will go through a wide range of noise impacts with Es/N0 ranging from -20 dB to 50 dB with a 10 dB increment step. To extract rich features, the CNN classifier applies 64 feature filters and therefore the first \ac{NN} sub-block outputs a three-dimensional 2${\times}$1024${\times}$64 feature matrix. To reduce the size of a feature matrix, downsampling functions such as MaxPool and AveragePool are applied. The full connection layer will resize the 2${\times}$1${\times}$64 input feature matrix to a 1${\times}$1${\times}\lambda$ output feature vector with $\lambda$ indicating the number of signal classes. In the end, the SoftMax layer computes the probability of each predicted signal class using the SoftMax function as
\begin{equation}\label{eq:SoftMax_function}
P_r(\psi_{i})=\frac{e^{\psi_{i}}}{\sum_{j=1}^{\lambda}e^{\psi_{j}}},
\end{equation}
where $\Psi=(\psi_1,\psi_2,...,\psi_{\lambda})\in\mathbb{R}^{\lambda}$ indicates the input feature vector to the SoftMax function and it includes $\lambda$ real numbers with the element index $i=1,2,...,\lambda$. The computation in \eqref{eq:SoftMax_function} ensures each output from the SoftMax is within the interval $[0,1]$ and the sum of each output equals one.

To find a classifier that works well for all the signal classes, cross entropy is computed as an indicator for the total loss as
\begin{equation}\label{eq:loss_function}
Loss=-\sum_{i=1}^{\lambda}P_r^{T}(\psi_{i}){\cdotp}\ln(P_r(\psi_{i})),
\end{equation}
where $P_r^{T}(\psi_{i})$ is the true probability that the $i^{th}$ input signal belongs to the $i^{th}$ signal class while $P_r(\psi_{i})$ is the predicted probability that the $i^{th}$ input signal belongs to the $i^{th}$ signal class. With the cross entropy calculation, the neural network can optimize its architecture via backward propagation using the Adam optimizer. {The maximum number of epochs is limited to 30 and the mini-batch size is 128. To fully extract features from a dataset, a learning rate of 0.01 is configured through the training.}

The signal pattern for each framework should be designed according to the pattern keys generated by the proposed chaotic sequence generator in section \ref{sec:signal_pattern_exchange}. {Ideally, the key, in other words the bandwidth compression factor $\alpha$, is continuous and therefore has an infinite number of values. This will advantageously show the robustness of our proposed framework in practice but the infinite number of values also complicate the evaluations of the proposed framework in simulations. Therefore, in this work, we use discrete values of $\alpha$ instead of using continuous values. The effect of a relatively small change of $\alpha$ has been studied in \cite{Tongyang_JIOT_WDS_2022} where the work showed that the narrower gap between adjacent values of $\alpha$, the lower classification accuracy is achieved. It is therefore expected that continuous values of $\alpha$ will lead to an infinite number of signal patterns, which will significantly decrease eavesdropping signal classification accuracy.}
 
The single-band WDS framework might be designed with the following SB signal pattern.
\begin{equation}
\left\{
  \begin{array}{l l}
    SB-OFDM &  \\
    SB-SEFDM &  \text{($\alpha$=0.95, 0.9, 0.85, 0.8, 0.75, 0.7)}
  \end{array} \right.,
\label{eq:SB_signal_pattern}\end{equation}
where the values in the bracket indicate the value of $\alpha$ for each signal class. The SB signal pattern has $\lambda$=7 signal classes and the \ac{BCF} gap between adjacent classes is $\Delta\alpha$=0.05. Each signal class has 2,000 OFDM/SEFDM symbols and there are overall 14,000 symbols for the SB signal pattern neural network training.

In terms of multi-band signals, this work will select sub-band \ac{BCF} $\beta$=0.9, 0.85, 0.8, which are a subset of the SB-SEFDM $\alpha$ pattern in \eqref{eq:SB_signal_pattern}. The bandwidth compression factor and the number of sub-carriers for each multi-band signal architecture is configured as the following.
\begin{equation}
\small
\left\{
  \begin{array}{l l}
    MB-1 &    \text{($\beta$=0.9, $N_B$=16)} \\
    MB-2 &    \text{($\beta$=0.85, $N_B$=16)} \\
    MB-3 &    \text{($\beta$=0.8, $N_B$=16)} 
  \end{array} \right..
\label{eq:MB_signal_pattern}\end{equation}

\begin{figure}[t!]
\begin{center}
\includegraphics[scale=0.24]{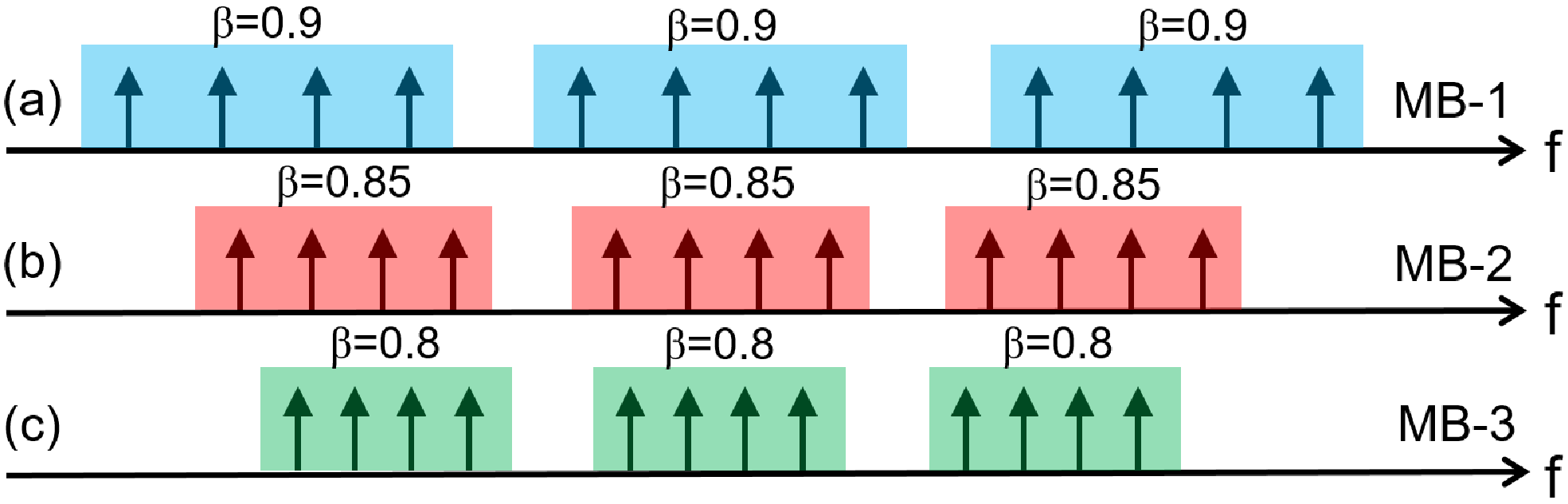}
\end{center}
\caption{Spectral packing characteristics for MB-SEFDM signal patterns. (a) MB-1. (b) MB-2. (c) MB-3.}
\label{Fig:MAMB_waveform_arch_all_MB}
\end{figure}

The MB-SEFDM signal pattern with $\lambda$=3 signal classes is designed in \eqref{eq:MB_signal_pattern} and illustrated in Fig. \ref{Fig:MAMB_waveform_arch_all_MB}, in which $\beta$=0.9, 0.85, 0.8 are allocated to Fig. \ref{Fig:MAMB_waveform_arch_all_MB}(a), Fig. \ref{Fig:MAMB_waveform_arch_all_MB}(b) and Fig. \ref{Fig:MAMB_waveform_arch_all_MB}(c), respectively. Each sub-band has the same number of sub-carriers $N_B$=16 but the variations of $\beta$ result in different spectral bandwidth. Each signal class has 2,000 symbols and there are overall 6,000 symbols for neural network training. 
\begin{equation}
\small
\left\{
  \begin{array}{l l}
    AMB-1 &    \text{($\beta$=0.9, $N_B$=16)} \\
    AMB-2 &    \text{($\beta$=0.85, $N_B$=17)} \\
    AMB-3 &    \text{($\beta$=0.8, $N_B$=18)} 
  \end{array} \right..
\label{eq:AMB_signal_pattern}\end{equation}

\begin{figure}[t!]
\begin{center}
\includegraphics[scale=0.24]{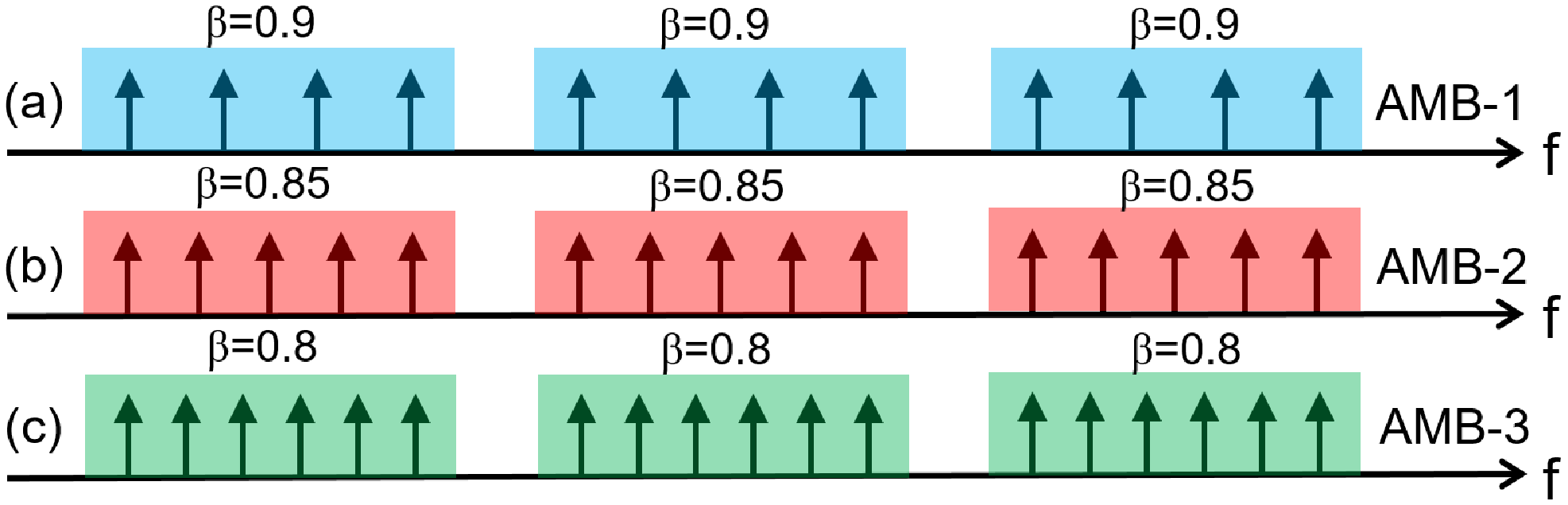}
\end{center}
\caption{Spectral packing characteristics for AMB-SEFDM signal patterns. (a) AMB-1. (b) AMB-2. (c) AMB-3.}
\label{Fig:MAMB_waveform_arch_all_AMB}
\end{figure}

The AMB-SEFDM signal pattern with $\lambda$=3 signal classes is designed in \eqref{eq:AMB_signal_pattern} and illustrated in Fig. \ref{Fig:MAMB_waveform_arch_all_AMB}. In order to have approximately similar occupied spectral bandwidth for each AMB signal, each sub-band in Fig. \ref{Fig:MAMB_waveform_arch_all_AMB}(a) with $\beta$=0.9 packs 16 sub-carriers, Fig. \ref{Fig:MAMB_waveform_arch_all_AMB}(b) and Fig. \ref{Fig:MAMB_waveform_arch_all_AMB}(c) should pack 17 and 18 sub-carriers, respectively. Each signal class has 2,000 symbols and there are overall 6,000 symbols for neural network training.
\begin{equation}
\small
\left\{
  \begin{array}{l l}
   MAMB-1 &    \text{($\beta$=0.9, 0.85, 0.8, $N_B$=16, 17, 18)} \\
   MAMB-2 &    \text{($\beta$=0.9, 0.85, 0.8, $N_B$=16, 17, 18)} \\
   MAMB-3 &    \text{($\beta$=0.9, 0.85, 0.8, $N_B$=16, 17, 18)} 
  \end{array} \right..
\label{eq:MAMB_signal_pattern}\end{equation}

\begin{figure}[t!]
\begin{center}
\includegraphics[scale=0.24]{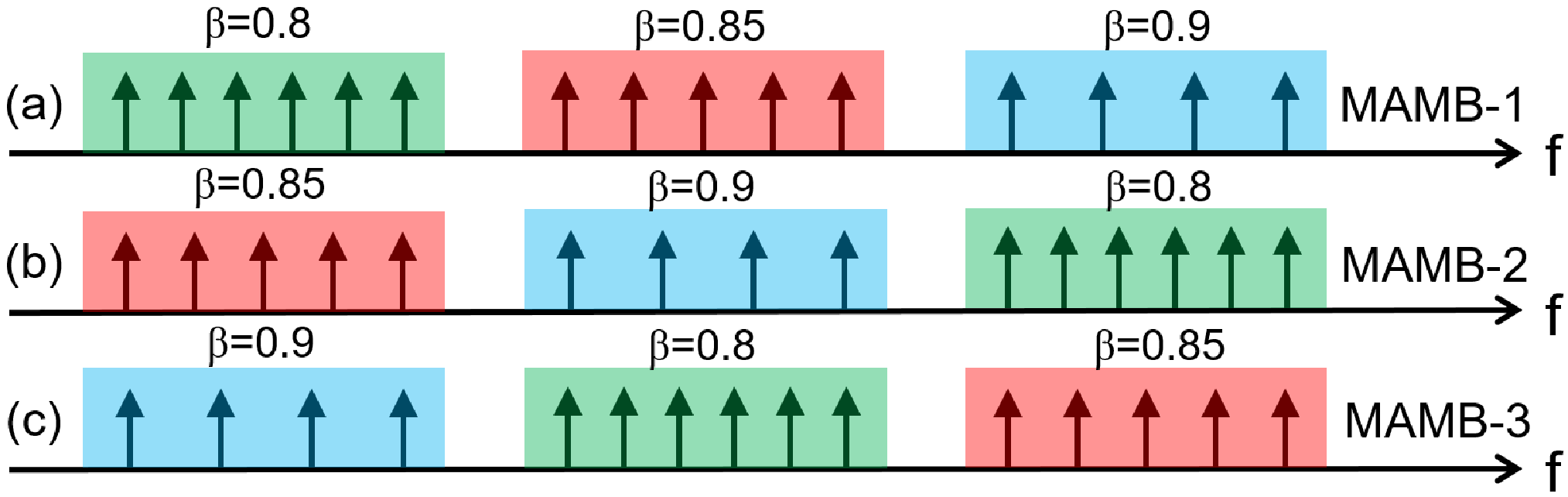}
\end{center}
\caption{Spectral packing characteristics for MAMB-SEFDM signal patterns. (a) MAMB-1. (b) MAMB-2. (c) MAMB-3.}
\label{Fig:MAMB_waveform_arch_all_MAMB}
\end{figure}

The MAMB-SEFDM signal pattern with $\lambda$=3 signal classes is designed in \eqref{eq:MAMB_signal_pattern} and illustrated in Fig. \ref{Fig:MAMB_waveform_arch_all_MAMB}. Similar to the AMB signal pattern, three different values of $\beta$ are employed. However, different $\beta$ would be mixed together in each signal class. Therefore, MAMB waveforms are similar to AMB waveforms in terms of occupied bandwidth but with different sub-band spectral features. The sub-band spectral ambiguity will cause misclassification at eavesdroppers. Each signal class has 2,000 symbols and there are overall 6,000 symbols for neural network training.

\subsection{Performance and Processing Complexity}

\begin{figure}[t!]
\begin{center}
\includegraphics[scale=0.6]{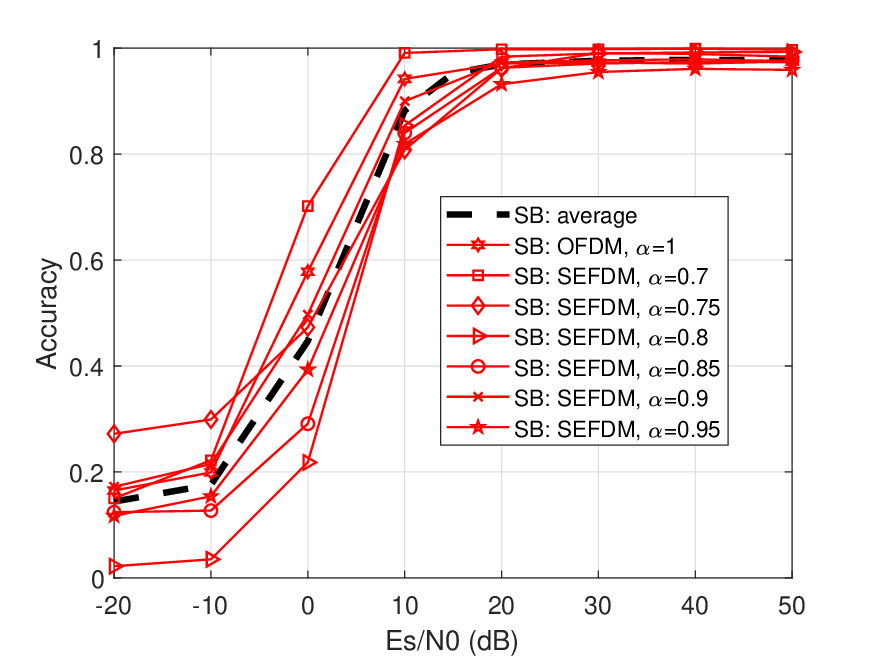}
\end{center}
\caption{Classification accuracy for SB based signal patterns and their average accuracy.}
\label{Fig:MAMB_SB_Type_II_accuracy}
\end{figure}

\begin{figure}[t!]
\begin{center}
\includegraphics[scale=0.6]{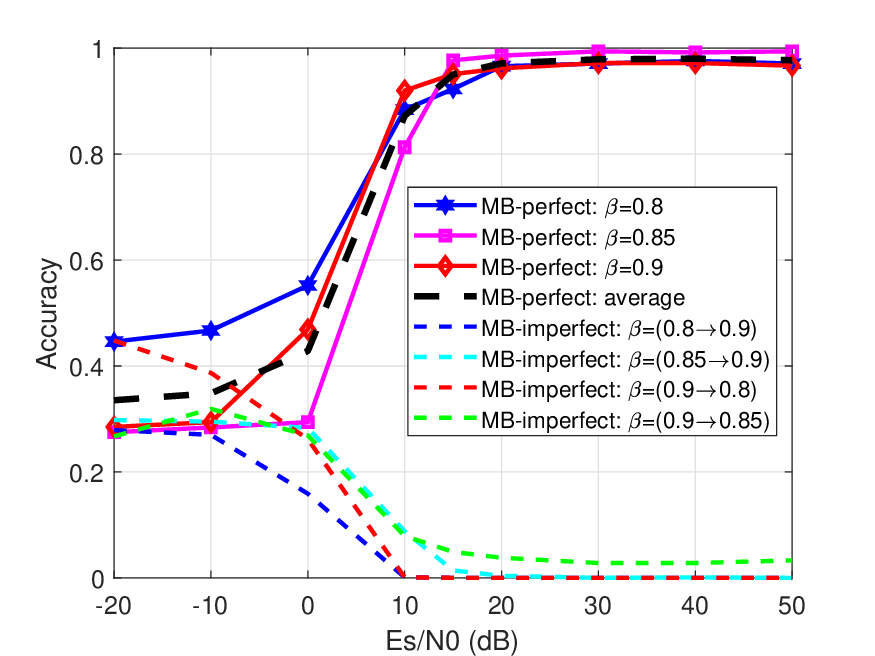}
\end{center}
\caption{Classification accuracy for MB based signal patterns and their average accuracy.}
\label{Fig:MAMB_MB_accuracy}
\end{figure}

{Due to the black-box learning mechanism of CNN, there is no analytical theory to justify the generality of the particular neural network in all security scenarios. To justify the feasibility of our trained CNN model in security analysis, we choose a benchmark for the reference. We firstly train a CNN model that can correctly classify the existing signal patterns. Then the CNN architecture will be re-trained for the newly proposed waveform scheme. In this case, we can have a fair justification that this particular CNN network is appropriate for the security analysis since the eavesdropping CNN model can eavesdrop conventional signals but it cannot identify the newly proposed signal patterns. Although an analytical justification is not available, the intensive training process for a CNN model results in time delay and will prevent eavesdropping in time-critical communications, which is a suitable application scenario that justifies the utility of our proposed framework.}

The classification accuracy of single-band signal patterns is shown as a benchmark in Fig. \ref{Fig:MAMB_SB_Type_II_accuracy}, in which all the signals can be identified at nearly 100\% accuracy rates with the increase of Es/N0. The classification accuracy results for multi-band signals are presented in Fig. \ref{Fig:MAMB_MB_accuracy}. Since there is no need for OFDM signals using a multi-band signal architecture, OFDM is not considered in the MB scenario. As expected from Fig. \ref{Fig:MAMB_MB_accuracy}, all the MB structured signals with perfect classification can converge to nearly 100\% accuracy at high Es/N0 regime. The imperfect classification for the target signal $\beta$=0.9 is also evaluated. The notation, $\beta=(\beta_0\rightarrow{\beta_1})$, indicates an imperfect classification from a signal class of $\beta_0$ to another signal class of $\beta_1$. The imperfect classification accuracy shows a complementary trend relative to its perfect accuracy. 

So far, both single-band and multi-band signal patterns are able to be identified by properly trained CNN classifiers. Compared to the single-band signal format, the multi-band signal architecture is a hardware-friendly signal format and its signal detection is implementable in hardware. However, they are both vulnerable to eavesdropping since eavesdroppers can apply deep learning to identify signals and employ proper algorithms to recover signals.

\begin{figure}[t!]
\begin{center}
\includegraphics[scale=0.6]{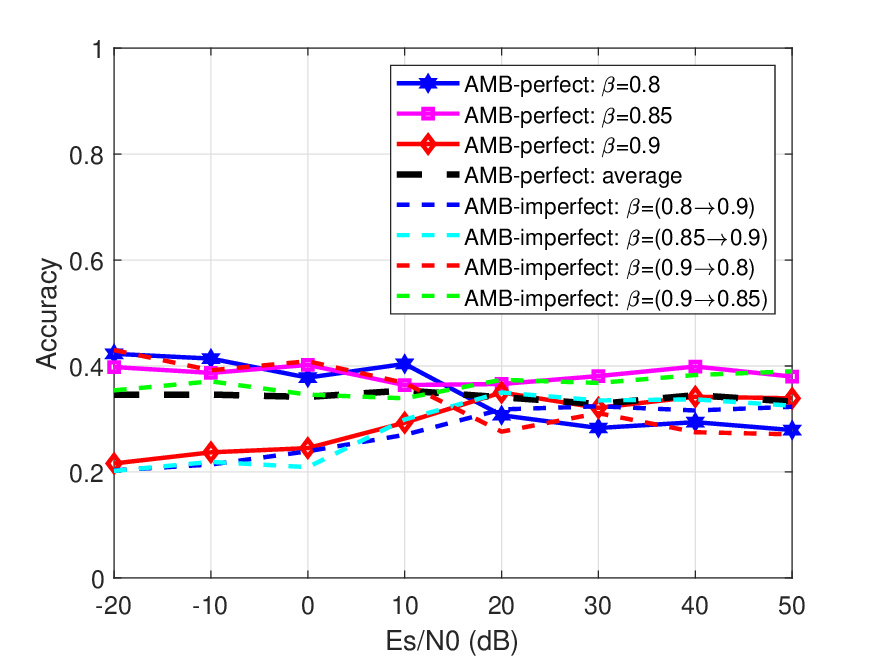}
\end{center}
\caption{Classification accuracy for AMB based signal patterns and their average accuracy.}
\label{Fig:MAMB_AMB_accuracy}
\end{figure}

The classification accuracy for the AMB signal pattern is investigated in Fig. \ref{Fig:MAMB_AMB_accuracy}. As usual, both perfect and imperfect classification results are presented. Unlike the complementary results observed from Fig. \ref{Fig:MAMB_MB_accuracy}, both perfect and imperfect classification accuracy rates are distributed around a static accuracy rate, 1/3. It is due to the fact that the three signal classes have strong feature similarity and each signal class would be equally classified into three signal classes resulting in the static 1/3 accuracy rate.

\begin{figure}[t!]
\begin{center}
\includegraphics[scale=0.6]{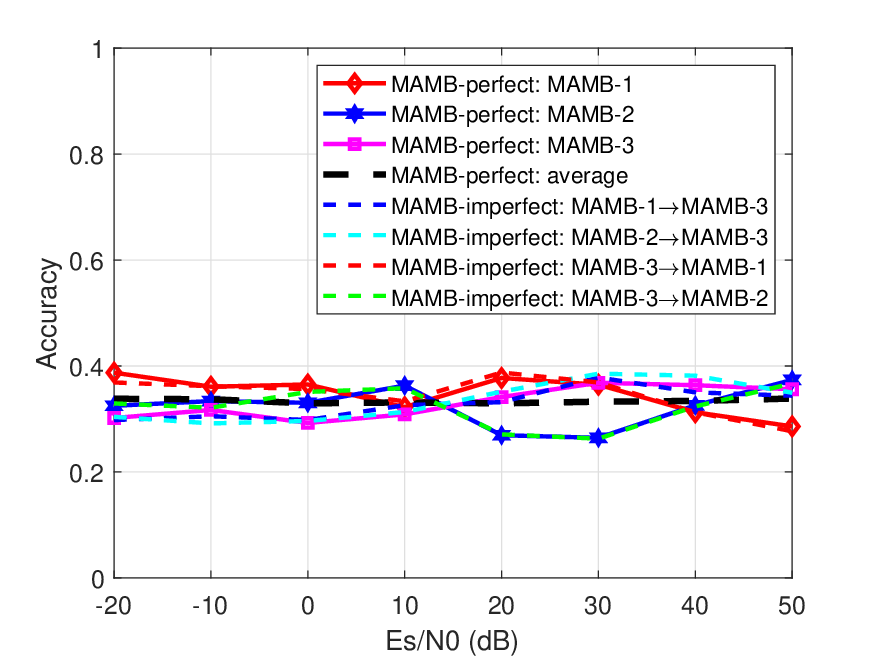}
\end{center}
\caption{Classification accuracy for MAMB based signal patterns and their average accuracy. The sub-band signal configuration follows the example in Table \ref{tab:MAMB_WDS_sub_band_architecture}.}
\label{Fig:MAMB_MAMB_accuracy}
\end{figure}

\begin{table}[t!]
\caption{MAMB-WDS sub-band architecture (an example used in this work)}
\centering
\begin{tabular}{cccc}
\hline \hline
Sub-band & MAMB-1  & MAMB-2 & MAMB-3 \\ 
Index &  $\beta$  &  $\beta$  &  $\beta$ \\ \hline 
0 & $0.90$ & $0.80$  & $0.85$\\ 
1 & $0.80$ & $0.90$  & $0.85$\\ 
2 & $0.85$ & $0.80$  & $0.90$\\ 
3 & $0.90$ & $0.90$  & $0.80$\\ 
4 & $0.90$ & $0.85$  & $0.90$\\ 
5 & $0.80$ & $0.90$  & $0.85$\\
6 & $0.85$ & $0.80$  & $0.90$\\ 
7 & $0.80$ & $0.80$  & $0.90$\\ 
8 & $0.90$ & $0.85$  & $0.80$\\ 
9 & $0.85$ & $0.85$  & $0.85$\\ 
10 & $0.90$ & $0.80$  & $0.80$\\ 
11 & $0.85$ & $0.90$  & $0.85$\\  
12 & $0.90$ & $0.85$  & $0.85$\\ 
13 & $0.80$ & $0.90$  & $0.80$\\ 
14 & $0.85$ & $0.80$  & $0.90$\\  
15 & $0.80$ & $0.85$  & $0.80$\\ \hline \hline
\label{tab:MAMB_WDS_sub_band_architecture}
\end{tabular}
\end{table}

To enhance further the ambiguity of classifying AMB signal patterns, the mixed signal pattern MAMB from \eqref{eq:MAMB_signal_pattern} is evaluated with classification accuracy presented in Fig. \ref{Fig:MAMB_MAMB_accuracy}, in which three mixed signal patterns are designed with the BCF characteristics in Table \ref{tab:MAMB_WDS_sub_band_architecture}. The 256 sub-carrier MAMB signal is divided into 16 sub-bands and each sub-band is allocated with a specific sub-band \ac{BCF} $\beta$ and an associated number of sub-carriers. In this case, three MAMB signal patterns effectively have the similar occupied spectral bandwidth. It should be noted that the combination pattern of sub-bands is flexible and Table \ref{tab:MAMB_WDS_sub_band_architecture} only shows an example. It is clearly seen from Fig. \ref{Fig:MAMB_MAMB_accuracy} that due to the randomness of each sub-band features, the enhanced ambiguity complicates MAMB signal classification resulting in the 1/3 accuracy rate at all Es/N0.

The results observed in Fig. \ref{Fig:MAMB_MAMB_accuracy} come with an assumption that the accuracy rate would be reduced further when more MAMB signal patterns are considered. The approximate accuracy rate to classify an arbitrary MAMB pattern is expressed in a mathematical model as
\begin{equation}\label{eq:accuracy_arbitrary_MAMB}
\psi=\frac{1}{\varpi},\ \ \ \ \varpi\in[1,2,3,...,b^{N/N_B}],
\end{equation}
where $\varpi$ indicates the number of MAMB signal classes, $b$ represents the number of \ac{BCF} candidates and $N/N_B$ indicates the number of sub-bands. Considering the example from Table \ref{tab:MAMB_WDS_sub_band_architecture}, it is clear that the example has $b$=3 due to $\beta$=0.9, 0.85, 0.8 and $N/N_B$=16 sub-bands. Therefore, the maximum number of MAMB signal classes is $\varpi=3^{16}$. In practice, the value of $\varpi$ would be infinite since the value of $b$ could be infinite due to continuous combinations of \ac{BCF}. In addition, the number of sub-bands $N/N_B$ is also flexible and the increase of the value will exponentially cut the classification accuracy rate.

\begin{figure}[t!]
\begin{center}
\includegraphics[scale=0.6]{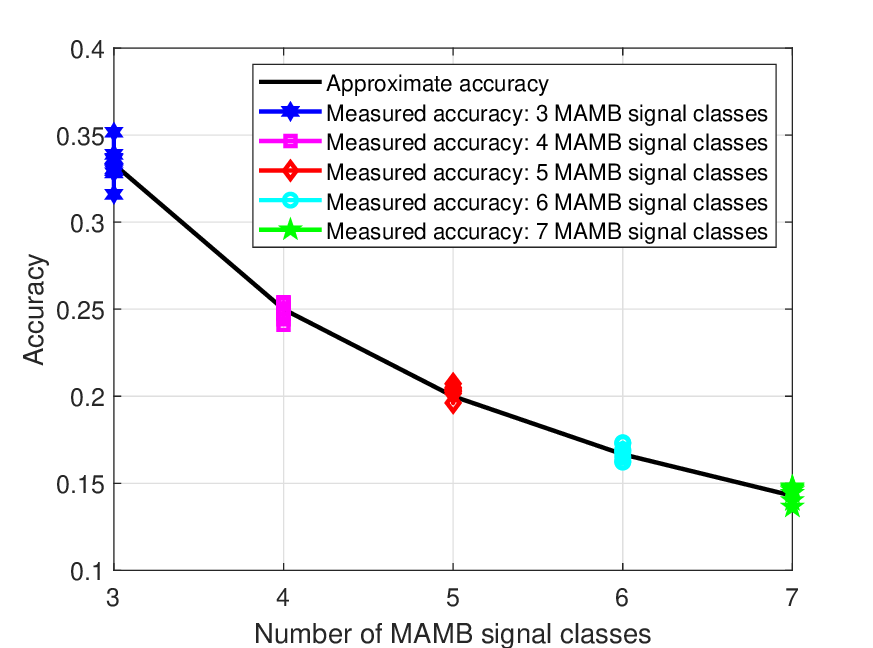}
\end{center}
\caption{Approximate accuracy and measured accuracy for MAMB signal patterns.}
\label{Fig:MAMB_analytical_vs_simulation_accuracy}
\end{figure}

\begin{figure}[t!]
\begin{center}
\includegraphics[scale=0.6]{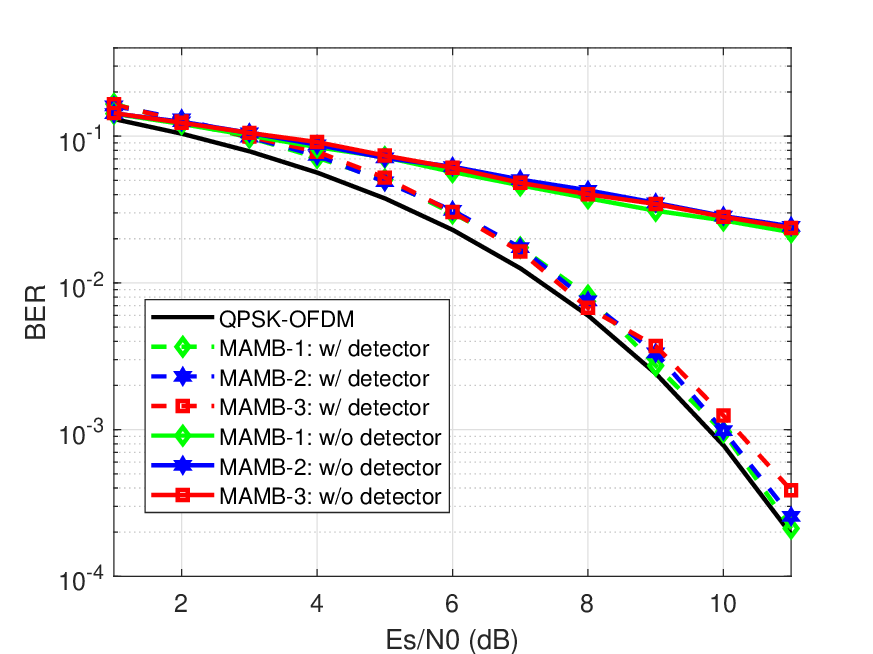}
\end{center}
\caption{BER performance for legitimate user MAMB signals with and without the uniquely designed SD detector.}
\label{Fig:MAMB_BER}
\end{figure}

Fig. \ref{Fig:MAMB_analytical_vs_simulation_accuracy} compares the approximate accuracy and measured accuracy for MAMB signal patterns with different number of signal classes.
Each signal pattern is evaluated ranging from Es/N0=-20 dB to Es/N0=50 dB with a 10 dB increment step. Therefore, each signal pattern will show eight evaluation points in Fig. \ref{Fig:MAMB_analytical_vs_simulation_accuracy}, in which it shows the reduction of classification accuracy with the increase number of signal classes. In addition, the measured accuracy reduction trajectory follows the accuracy approximation in \eqref{eq:accuracy_arbitrary_MAMB} where the accuracy rate drops by 57\% from three signal classes to seven signal classes.

In addition to the robustness evaluations of the WDS framework to prevent eavesdropping, Fig. \ref{Fig:MAMB_BER} shows communication reliability at legitimate users as well. The MAMB signal pattern, with three signal classes, is selected for BER testing. The legitimate user will use pre-known pattern information to detect signals. It reveals that without a proper signal detector, where \ac{MF} is applied, all the MAMB signal classes cannot be decoded resulting in high BER results. On the other hand, with the help of a uniquely designed detector, where the SD architecture from section \ref{subsec:principle_signal_detection} is applied for each signal sub-band, all the signal classes are detectable with similar performance to QPSK-OFDM. Based on the results in Fig. \ref{Fig:MAMB_BER}, it is inferred that even signals are correctly identified by eavesdroppers, they cannot decode signals properly when the uniquely designed SD detector is not known in advance.

\begin{figure}[t!]
\begin{center}
\includegraphics[scale=0.58]{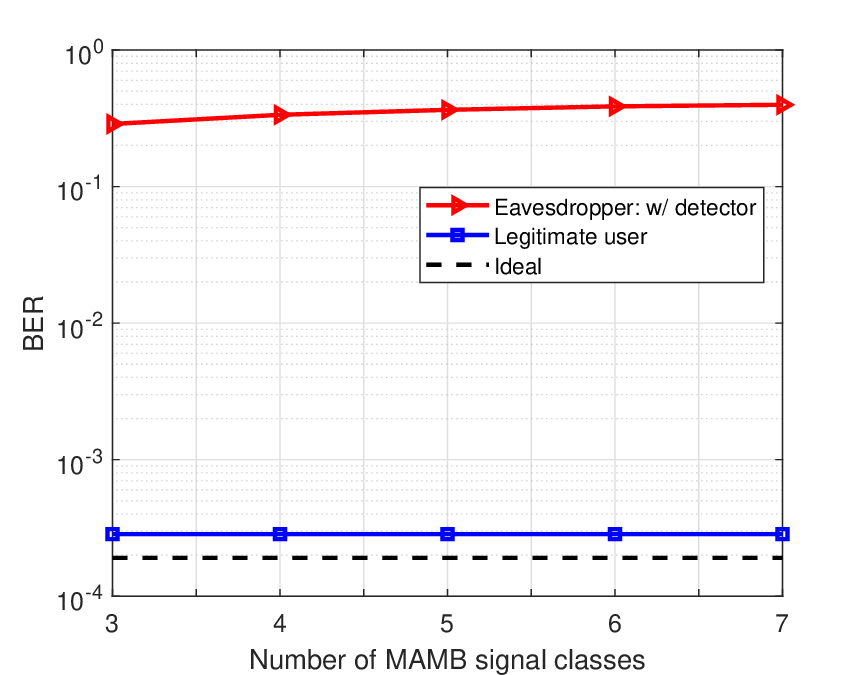}
\end{center}
\caption{BER performance for eavesdroppers with knowledge of the uniquely designed SD detector.}
\label{Fig:BER_MAMB_eav_legitimate_ideal}
\end{figure}

To explore the eavesdropping capability on MAMB signals, Fig. \ref{Fig:BER_MAMB_eav_legitimate_ideal} shows that the eavesdropping performance approaches a flat BER curve even the multiband SD detector is employed indicating a failure of eavesdropping. Based on the results in Fig. \ref{Fig:BER_MAMB_eav_legitimate_ideal}, it is inferred that even when the uniquely designed SD detector is known by eavesdroppers in advance, they cannot decode signals properly because signals are not correctly identified by eavesdroppers, which further enhances the physical layer communication security.
\begin{table*}[t]
\centering
\caption{ Signal Processing Complexity Analysis (uplink channel from Alice to Bob)}
\begin{tabular}{ | c | c | c || c | c | c | } \hline

$\mathbf{Processing}$ & WDS(Alice)  & WDS(Bob) & multi-band WDS(Alice)  & multi-band WDS(Bob) & WDS/multi-band WDS(Eve)  \\ 
           &  User  & Base Station    & User  & Base Station & Eavesdropper \\ \hline \hline

$\mathbf{Tx}$ & IFFT(single) & - & IFFT(multiple) & - & - \\ \hline 
   &     &   FFT(single) &       &            FFT(multiple)          & FFT(single/multiple)\\ 
$\mathbf{Rx}$ &  - &  Signal Detection  &   -       & Signal Detection  & Signal Classification\\ 
   &      &      &         &       & Signal Detection\\ \hline

\end{tabular}
\label{tab:MAMB_WDS_signal_processing_complexity_analysis}
\end{table*}

The signal processing complexity for WDS and multi-band WDS frameworks is compared in Table \ref{tab:MAMB_WDS_signal_processing_complexity_analysis}. The pattern key generation is one-time processing and is not taken into account. At the transmitter (Tx), the traditional WDS requires single IFFT while the proposed multi-band WDS requires multiple IFFTs. In terms of receiver side (Rx), both frameworks are for uplink channel communications where complex signal processing is at energy consumption insensitive base stations. {Therefore, signal detection complexity is not the limitation to our proposed security framework. In summary, our proposed framework significantly reduces power consumption in Fig. \ref{Fig:MAMB_power_consumption} due to the reduced hardware utilization analysed in Table \ref{tab:WDS_power_complexity_analysis}.}

\begin{figure}[t!]
\begin{center}
\includegraphics[scale=0.47]{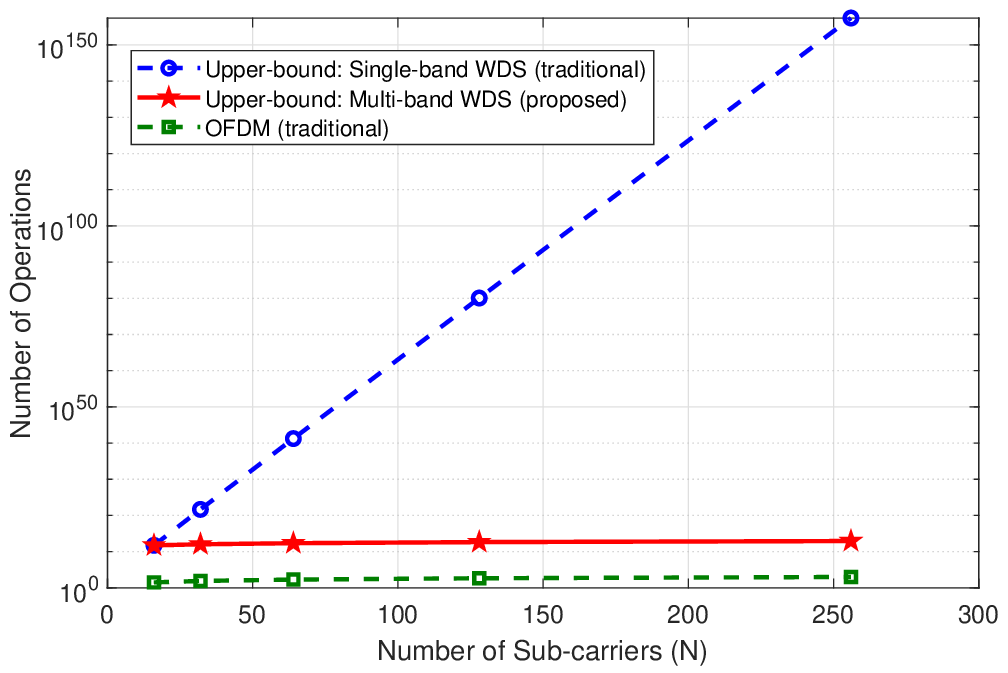}
\end{center}
\caption{{The upper-bound of signal detection complexity in the number of real-valued multiplication operations.}}
\label{Fig:AI_complexity_SD_vs_FFT_vs_MultiSD}
\end{figure}

{Compared to traditional OFDM, our proposed waveform framework has increased spectral efficiency and higher data rate in a given bandwidth. It is because our proposed waveform framework can compress occupied spectral bandwidth and generate non-orthogonal waveforms. As a result, in a given spectral bandwidth, we can pack more sub-carriers for carrying data, leading to an increased data rate. The obvious limitation of our proposed approach, compared to OFDM, is the increased signal processing complexity at the receiver side, because the system requires complex signal detection algorithms to decode signals at legitimate users. We have implemented a similar signal detector in our previous work \cite{TongyangICT2013}, which verifies that the data rate can be enhanced using optimized digital circuit design. To evaluate signal detection complexity, real-valued multiplication operations are considered. Since \ac{SD} has variable computational complexity related to the level of noise power, this work will evaluate the upper-bound complexity. For traditional OFDM based systems, signal detection relies on \ac{MF}, which is the \ac{FFT} operation with the computational complexity of $(N/2)log2(N)$ multiplications. For the traditional single-band WDS framework, a single \ac{SD} detector is required with the upper bound complexity of $\sum_{n=1}^{2N}2^n[2n+1]$. For our newly proposed multi-band WDS framework, its signal detection has upper bound complexity of $\frac{N}{N_B}\sum_{n=1}^{2N_B}2^n[2n+1]$. Fig. \ref{Fig:AI_complexity_SD_vs_FFT_vs_MultiSD} clearly shows the complexity difference for the three waveform schemes. It is obvious that the newly proposed multi-band WDS framework has higher detection complexity compared to the traditional OFDM scheme but our proposal has significant complexity reduction compared to the single-band WDS framework.}

\begin{figure}[t]
\begin{center}
\includegraphics[scale=0.58]{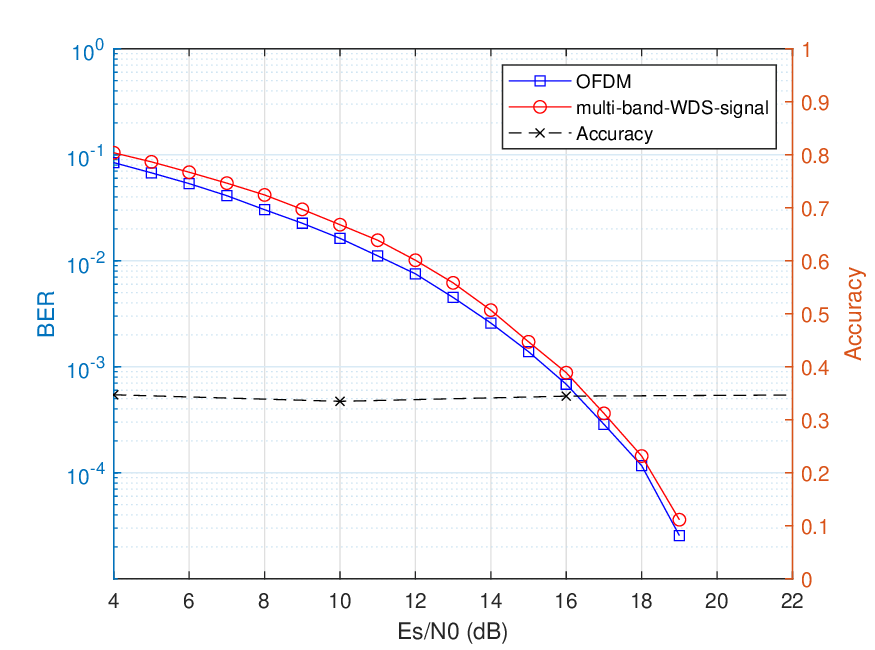}
\end{center}
\caption{{BER performance and classification accuracy evaluations in multipath fading channels.}}
\label{Fig:TWC_2023_2nd_review_BER_channel_A}
\end{figure}

{It is noted that this work obtains the security enhancement capability using non-orthogonal signal waveform ambiguity rather than relying on channel variations. Our previous work \cite{Tongyang_JIOT_WDS_2022} has verified that channels have minimal effects on classification since eavesdroppers fail to distinguish signals in both AWGN and wireless channels. For further information on the effect of channels, previous studies in \cite{TongyangTVT2017, TongyangJLT2016} have verified the feasibility of the non-orthogonal signals in practical experiment. To provide a comprehensive evaluation, we test our proposed signal scheme under the multipath fading channel model \cite{TongyangTVT2017,OFDM_robust2005,Channel_Yunsi_CL_2020} where each path is configured to experience Rayleigh fading. In Fig. \ref{Fig:TWC_2023_2nd_review_BER_channel_A}, our proposed multi-band WDS signal exhibits close BER performance to OFDM in multipath fading scenarios, suggesting that the proposed non-orthogonal signal can provide good BER performance in fading channels. Fig. \ref{Fig:TWC_2023_2nd_review_BER_channel_A} also demonstrates the classification accuracy at eavesdropper under multipath fading channels. It is evident that the accuracy is not obviously affected by channels, because the classification relies on waveform spectral ambiguity rather than channel variations.}

\begin{figure}[t]
\begin{center}
\includegraphics[scale=0.56]{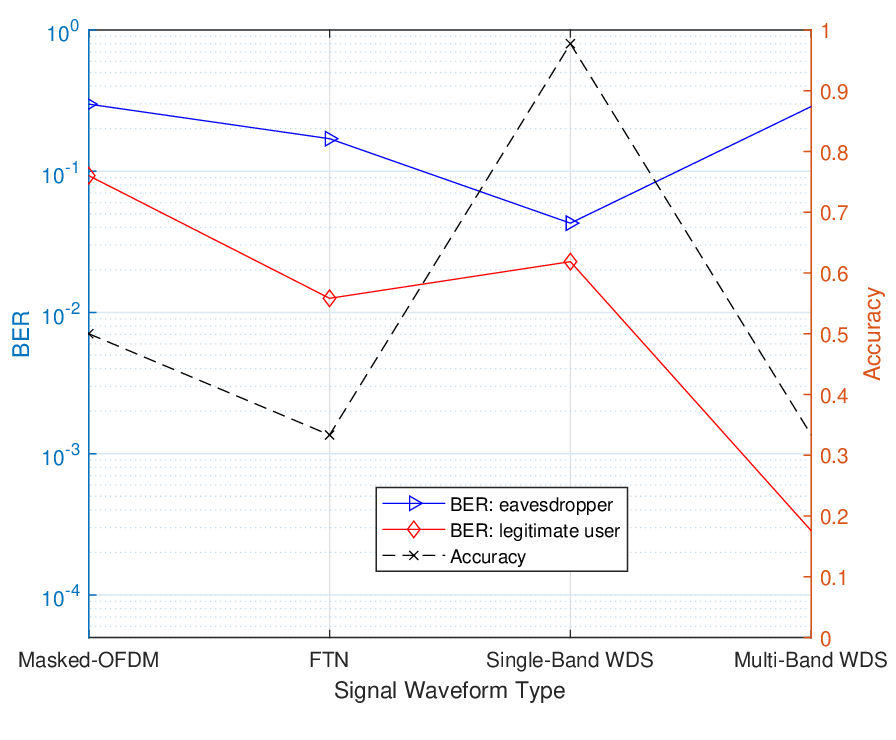}
\end{center}
\caption{{Comparison with other waveform schemes in terms of classification accuracy, BER at legitimate users and eavesdroppers.}}
\label{Fig:TWC_2023_2nd_review_compare}
\end{figure}

{We include the Masked-OFDM technique \cite{MOFDM_PIMRC2009}, the FTN-based technique \cite{PLS_FTN_ICCE_2017}, and the single-band SEFDM-based technique \cite{Tongyang_JIOT_WDS_2022} for the comparison in this paper. All of these waveforms are incorporated into multicarrier formats similar to the multi-band WDS framework, and they all exhibit higher spectral efficiency compared to OFDM. The classification accuracy results reveal that the eavesdropper achieves the lowest accuracy (therefore better security performance) by our proposed multi-band WDS signal, while the eavesdropper attains the highest accuracy by the single-band WDS signal. For the other two signals, their accuracy is similar to that of the multi-band WDS signal. Concerning BER performance, the aim is to develop a framework that increases the eavesdropper's BER while simultaneously reducing the legitimate user's BER. Based on the results presented in Fig. \ref{Fig:TWC_2023_2nd_review_compare}, it is evident that our proposed multi-band WDS framework can meet both requirements while all other waveform candidates degrade both legitimate user and eavesdropper BER performance.}

\section{Conclusion} \label{sec:conclusion}

This work investigated a multi-band waveform-defined security (WDS) framework, which avoids CSI at transmitters and can be jointly used with traditional PLS techniques. An adaptive multi-band WDS scheme is able to confuse eavesdropping signal identification since the designed signals occupy the same spectral bandwidth while their sub-band spectral characteristics are variable and unknown by eavesdroppers. With adaptive adjustment of each sub-band spectral feature, the eavesdropping accuracy drops to 33\% when only three sub-band signal classes are taken into account. It is noted that spectral features for each sub-band are determined by sub-carrier packing patterns, which theoretically have an infinite number of combinations due to the continuous variations of the packing schemes. Therefore, the potentially infinite combinations of WDS patterns can efficiently prevent brute-force eavesdropping. An accuracy approximation model is derived to reveal that the eavesdropping accuracy will drop further when the number of feature combinations increases. Results show a nearly 57\% accuracy drop when the number of combinations goes from three to seven. Signal BER performance is also evaluated and results show nearly perfect signal recovery with lower complexity relative to traditional PLS approaches.

\bibliographystyle{IEEEtran}
\bibliography{TWC_Ref}

\end{document}